\documentclass[a4paper,fleqn,usenatbib,useAMS]{mnras}

\usepackage{graphicx}	
\usepackage{amsmath}	
\usepackage{amssymb}	
\usepackage{multicol}        
\usepackage{bm}		
\usepackage{pdflscape}	

\newcommand{\kms}{\,km\,s$^{-1}$} 

\defcitealias{Coccato_2020}{Paper~I}
\defcitealias{Thomas_2011}{TMJ}

\usepackage[T1]{fontenc}
\usepackage{ae,aecompl}

\usepackage{newtxtext,newtxmath}

\title[Formation of S0s in extreme environments II]{Formation of S0s in extreme environments II: the star-formation histories of bulges, discs and lenses}

\author[E. J. Johnston \textit{et al}]{Evelyn J. Johnston$^{1,2}$\thanks{Contact e-mail: \href{mailto:evelynjohnston.astro@gmail.com}{evelynjohnston.astro@gmail.com}}, Alfonso Arag\'on-Salamanca$^{3}$, Amelia Fraser-McKelvie$^{3,4}$, 
\newauthor  Michael Merrifield$^{3}$, Boris H\" au\ss ler$^{5}$, Lodovico Coccato$^{6}$, Yara Jaff\'e$^{7}$, Ariana Cortesi$^{8,9}$, 
\newauthor  Ana Chies-Santos$^{10}$, Bruno Rodr\'iguez Del Pino$^{11}$, \& Yun-Kyeong Sheen$^{12}$.
\\
$^{1}$Institute of Astrophysics, Pontificia Universidad Cat\'olica de Chile, Av.~Vicu\~na Mackenna 4860, 7820436 Macul, Santiago, Chile\\
$^{2}$ N\'ucleo de Astronom\'ia de la Facultad de Ingenier\'ia y Ciencias, Universidad Diego Portales, Av. Ej\'ercito Libertador 441, Santiago, Chile. \\
$^{3}$ School of Physics and Astronomy, University of Nottingham, University Park, Nottingham NG7 2RD, UK\\
$^{4}$ International Centre for Radio Astronomy Research, The University of Western Australia, 35 Stirling Hw, 6009 Crawley, WA, Australia\\
$^{5}$ European Southern Observatory, Alonso de C\'ordova 3107, Vitacura, Santiago, Chile \\
$^{6}$ European Southern Observatory, Karl-Schwarzchild-str., 2, D-85748 Garching b. Munchen, Germany\\
$^{7}$ Instituto de Física y Astronom\'ia, Universidad de Valpara\'iso, Gran Bretana 1111, Valpara\'iso, Chile\\
$^{8}$ Instituto de Astronomia, Geof\'isica e Ci\^encias Atmosf\'ericas (IAG), Universidade de S\'ao Paulo (USP), R. do Mat\~ao 1226, 05508-090 S\~ao Paulo, Brazil\\ 
$^{9}$ Observatorio de Valongo, Universidade Federal do Rio de Janeiro, Ladeira do Pedro Antonio, 43, Centro, Rio de Janeiro - RJ 20080-090, Brazil \\
$^{10}$ Departamento de Astronomia, Universidade Federal do Rio Grande do Sul, Av. Bento Gon\c{c}alves 9500, Porto Alegre 91501-970, RS, Brazil \\
$^{11}$ Centro de Astrobiolog\'ia (CSIC-INTA), Torrej\'on de Ardoz, Madrid, Spain \\
$^{12}$ Korea Astronomy and Space Science Institute, 776 Daedeokdae-ro, Yuseong-gu, Daejeon 34055, Republic of Korea \\
}

\pubyear{2020?}

\begin{document}
\label{firstpage}
\pagerange{\pageref{firstpage}--\pageref{lastpage}}
\maketitle

\begin{abstract}
Different processes have been proposed to explain the formation of S0s, including mergers, disc instabilities and quenched spirals. These processes are expected to dominate in different environments, and thus leave characteristic footprints in the kinematics and stellar populations of the individual components within the galaxies.  New techniques  enable us to cleanly disentangle the kinematics and stellar populations of these components in IFU observations. In this paper, we use \textsc{buddi} to spectroscopically extract the light from the bulge, disc and lens components within a sample of 8 S0 galaxies in extreme environments observed with MUSE. While the spectra of bulges and discs in S0 galaxies have been separated before, this work is the first to isolate the spectra of lenses. Stellar populations analysis revealed that the  bulges and lenses have generally similar or higher metallicities than the discs, and the $\alpha$-enhancement of the bulges and discs are correlated, while those of the lenses are completely unconnected to either component. We conclude that the majority of the mass in these galaxies was built up early in the lifetime of the galaxy, with the bulges and discs forming from the same material through dissipational processes at high redshift. The lenses, on the other hand, formed over independent timescales at more random times within the lifetime of the galaxy,  possibly from evolved bars. The younger stellar populations and asymmetric features seen in the field S0s may indicate that these galaxies have been affected more by minor mergers than the cluster galaxies.

\end{abstract}





\section{Introduction}
The lenticular, or S0, classification is often used for galaxies whose morphology is intermediate between ellipticals and spirals, such that they display the same discy structure seen in spirals but with the redder colours and old stellar populations typically found in ellipticals. S0s are now considered, in most cases, to be evolved spirals in which the star formation has been truncated and the spiral arms have faded \citep[e.g.][]{Bedregal_2006, Moran_2007, Laurikainen_2010, Cappellari_2011, Prochaska_2011, Kormendy_2012, Johnston_2014}. Further evidence for this scenario lies in the morphology-density relation, which has shown that the frequency of S0s increases with increasing environmental density while the frequency of spirals decreases proportionally \citep{Dressler_1980, Dressler_1997, Cappellari_2011}. Many mechanisms have been proposed to explain this transformation, such as the gas being stripped out via ram-pressure stripping \citep{Gunn_1972}, harassment \citep{Moore_1998} or strangulation \citep{Larson_1980}, or used up in a rapid star formation event following a minor merger \citep{Ponman_1994, Arnold_2011} or successive minor mergers \citep{Bekki_2011}. Furthermore, simulations have  shown that major mergers can create a bulge around which a disc is able to form to ultimately create an S0 galaxy \citep{Spitzer_1951, Tapia_2017, Diaz_2018, Eliche_2018, Mendez_2018}, or even a spiral galaxy \citep{Fabricius_2014}. All of these processes require an external influence, such as an interaction with a neighbouring galaxy or the intracluster medium, but the existence of S0s in isolated environments indicates that other evolutionary processes must be possible. For example, \citet{Eliche_2013} proposed that S0s could simply be remnants of spirals after the gas reservoirs were exhausted and the spirals arms have faded, while \citet{Saha_2018} suggested that low-mass S0s may have formed through disc instabilities and fragmentation.

Many studies have attempted to determine the conditions under which these different processes are more dominant in transforming a spiral galaxy into an S0, such as dependencies on the mass of the galaxy and the local environment. For example, evidence of a mass dependence in the formation of S0s has been found by \citet{Fraser_2018b}. They studied the stellar populations within the bulges and discs of a sample of S0s from the MaNGA survey, and concluded that S0s with masses $>10^{10}M_\odot$ transformed via morphological or inside-out quenching, while lower mass counterparts have undergone bulge rejuvenation or disc fading. Another study by \citet{Dominguez_2020} concluded that S0s with masses $>10^{10}M_\odot$ have formed mainly through mergers, while those with $<10^{10}M_\odot$ are closer to spirals in their formation, and thus in-situ star formation has played a more significant role than mergers. 

The first paper in this series \citep[][hereafter Paper~I]{Coccato_2020}  investigated the effect of both the mass and environment on the formation of S0s through studies of their kinematics and global stellar populations. The kinematics revealed that the cluster galaxies are more rotationally supported while the isolated galaxies are more dispersion supported, thus suggesting different formation mechanisms, i.e. through gas removal (e.g. starvation, ram-pressure stripping) and minor mergers respectively. Another study of the kinematics of S0 galaxies in the SAMI Survey \citep{Green_2018} by \citet{Deeley_2020} also found the same result. The integrated stellar populations studied in \citetalias{Coccato_2020}, on the other hand, revealed no significant differences as a function of environment, but a tentative correlation was found with the stellar mass of the galaxies such that more massive galaxies contained older and more-metal-rich stellar populations. However, this stellar populations analysis was limited to the global properties within $1R_e$, where the bulge dominates the light, and so a more detailed stellar populations analysis across the galaxies is therefore needed to better understand this trend.

The different transformation processes would leave characteristic signatures within the physical structure and stellar populations of the different components within the  final galaxy, and so by studying these properties we can better understand how S0s formed. S0s were historically seen as featureless discs with, in many cases, a spheroidal bulge dominating the light at the core. However, more recent studies have found that their structures are not so straightforward. For example, bulges are no longer seen as simple spheroids, and instead can be classified as classical or pseudo-bulges based on the steepness of their surface brightness profiles and whether they are dissipationally or rotationally supported structures \citep{Kormendy_2004}.

Additionally, S0s are now known to host many distinct substructures, such as bars, ovals and lenses \citep[e.g.][]{Laurikainen_2005, Laurikainen_2009}. \citet{Kormendy_1979} defined a lens as an independent component with distinct surface brightness profiles, which are morphologically intermediate between bulges and discs with their flatter surface brightness profiles and sharp edges. Ovals have similar surface brightness profiles to lenses, but are more elongated, while bars display even higher ellipticities and can be detected through distortions in the kinematics across the galaxy \citep[e.g.][]{Bureau_1999,Merrifield_1999, Gadotti_2005}.  The true nature of lenses and ovals is still uncertain, although many current theories identify them as either remnants of dissolved bars \citep{Kormendy_1979} or artefacts from major mergers that are associated to stellar halos or embedded inner discs \citep{Eliche_2018}. It was originally thought that lenses and ovals were related to the host galaxy morphology, where lenses are found in galaxies of S0 and S0/a morphologies while ovals reside in later type disc galaxies. However, \citet{Laurikainen_2006} found that S0s can host both lenses and ovals, reflecting that the distinction between these classifications based on the galaxy morphology is not so straightforward.

While lenses have been seen in galaxies of all discy morphologies, they appear most frequently in S0s \citep{Laurikainen_2009}, both barred and unbarred \citep{Laurikainen_2005,Laurikainen_2007}. The fraction of S0s reported to display lenses and ovals varies from $\sim15\%$ \citep{Nair_2010} up to $97\%$ \citep{Laurikainen_2009}. 
\citet{Buta_2010} proposed that lenses and ovals are the remnants of bars that have become disrupted during the transformation from spirals to S0, and thus are created through secular evolution. As further evidence towards this scenario, the frequency of lenses from spirals  to S0s increases proportionally to the decrease in the frequency of bars between the same two morphologies \citep[e.g.][]{Laurikainen_2009}. Additionally, bars in S0s have been seen to be in general more evolved than in spirals, being longer, more massive and more often with ansae morphologies \citep[bars with bright enhancements at each end;][]{Elmegreen_1985,Laurikainen_2007, Martinez_2007, Diaz_2016}. Gas infall has also been proposed by \citet{Pfenniger_1990} and \citet{Laurikainen_2010} in connection with the secular evolution theory, whereas \citet{Eliche_2018} showed that major mergers can also create these structures using data from the dissipative merger simulations of the GalMer database \citep{Chilingarian_2010}.

Many imaging surveys have been carried out to study the structural parameters of the different structures within S0s. For simplicity, the larger, more statistical surveys have focussed on modelling the galaxies using a single or double light profile, where the single profile was usually a S\'ersic profile and gives information on the global structure of the galaxy while double profiles usually model an exponential disc with either a de~Vaucouleurs or S\'ersic bulge \citep[e.g.][]{Simard_2011, Mendel_2014, Vika_2014, Head_2014, Bottrell_2019}.  However, failures to account for all the components within a galaxy can affect the results. For example, excess light from a lens or bar may be attributed to the bulge, thus skewing the B/T light ratio to higher values and affecting the measurements for the effective radius and S\'ersic profile for that component. Additionally, \citet{Head_2015} showed that discs are not always perfectly exponential, but instead show a wide range of S\'ersic indices. As a result, limiting the fit to the discs to this profile may also result in light in the inner regions of the galaxy being misattributed to the bulge or disc. As a result, a few recent studies have expanded their fits beyond such simple models for their galaxies. For example, the Near-IR S0 Galaxy Survey \citep[NIRS0S,][]{Laurikainen_2011}  modelled $K_s$-band images of S0s and classified their morphologies in terms of bulges, discs, bars, lenses, barlenses etc, and  \citet{Head_2015} modelled $g$, $r$ and $i$-band images of galaxies in the Coma Cluster with up to three S\'ersic components to take into account lenses, bars and breaks or truncations in the discs. This latter study found that the fraction of pure S\'ersic profiles increases towards the centre of the cluster while the fraction of exponential discs is higher in the outskirts, but found no significant variation in the fraction of multi-S\'ersic systems with radius within the cluster.

In order to better understand the different formation scenarios of bulges, discs and lenses in S0s, and thus their roles in the transformation from spiral to S0, one needs to spectroscopically separate the light from each of these components. With the development of Integral Field Unit (IFU) spectrographs with wide fields of view and high spatial resolution, it is now possible to model the light profile of a galaxy in each image slice in order to cleanly extract the spectrum for each component included in the model with minimal contamination from the superposition of light from the other structures. One approach to carry out this modelling and extraction of spectra is Bulge--Disc Decomposition of IFU data \citep[\textsc{buddi},][]{Johnston_2017}. In this paper, we apply this novel approach to a sample of 8 S0s observed with the Multi-Unit Spectroscopic Explorer \citep[MUSE,][]{Bacon_2010} at the Very Large Telescope (VLT) in order to identify the physical structures within these galaxies and derive estimates of their stellar populations and star formation histories from their spectra. While several studies in the literature have extracted spectroscopic stellar populations for bulges and discs of S0s \citep[e.g.][]{Silchenko_2012, Johnston_2012, Johnston_2014, Coccato_2017, Tabor_2017, Tabor_2019, Mendez_2019a, Mendez_2019b, Oh_2020}, this work achieves the first such analysis of lenses. Through this analysis, we  use the combined photometric and spectroscopic information to build up a better understanding of how each component has formed and evolved, and their role in shaping the galaxies we see today.

This paper is laid out as follows: Section~\ref{sec:DR} describes the sample selection and data reduction;  Section~\ref{sec:radial_stellar_pops} studies the the stellar populations across each galaxy; Section~\ref{sec:buddi} outlines the spectroscopic decomposition technique used to cleanly model and extract the spectra of each component; Section~\ref{sec:overview_fits} discusses the components included in the models; Section~\ref{sec:stellar_pops}  presents the stellar populations analysis for each component through line strength analysis and full spectral fitting, and finally Section~\ref{sec:Discussion} discusses the results and presents the conclusions.  Throughout this paper we assume a Hubble constant of H$_\text{o}=70\pm1.3$~km~s$^{-1}$~Mpc$^{-1}$ \citep{Carroll_1992}.


\section{Data sample and reduction}\label{sec:DR}
This study uses observations of a sample of 8 S0 galaxies that were observed with MUSE in Wide-Field Mode without adaptive optics as part of ESO Programme 096.B-0325 (PI: Jaff\'e). The sample consists of four galaxies from the Centaurus Cluster and four galaxies identified as lying in isolated environments according to the 2MASS Isolated Galaxies Catalogue \citep[2MIG,][]{Karachentseva_2010}. The galaxies were selected to have  morphological Hubble types of $-3.5 < \text{T} < -0.5$ from LEDA,  clear bulge+disc structure with no evidence of bars or tidal disruptions, and to have inclinations above 40$^{\circ}$ to reduce the chances of including misclassified ellipticals in the sample. The Centaurus Cluster galaxies lie at a distance of 48.7~Mpc ($z\sim0.011$) while the isolated galaxies have distances of $60-160$~Mpc ($z\sim0.014-0.037$), and all galaxies have masses within the range $\text{log}(M/M_\odot)=10.6-11.6$. Full details of the observations and data reduction can be found in \citetalias{Coccato_2020}.


\section{Radial Stellar Populations}\label{sec:radial_stellar_pops}

In \citetalias{Coccato_2020}, the global mass-weighted stellar populations were measured within $1~R_e$ for each galaxy. Due to the small sample, no clear differences were detected as a function of environment. However, a dependence on mass was detected, where higher mass galaxies tend to have higher ages and metallicities, and built up $50\%$ of their total mass earlier on in their lifetimes. Similar trends were found for ETGs in \citet{McDermid_2015}, and for bulges and discs of S0s in \citet{Fraser_2018b}. 

In this section, the stellar populations across each galaxy will be studied in more detail. As described in \citetalias{Coccato_2020}, six of the galaxies in this sample contain H$\alpha$ emission, either concentrated in the core or distributed throughout the galaxy. The results presented in this paper focus on the analysis of the underlying stellar populations as measured from the stellar continuum and absorption features.

\begin{figure*}
 \includegraphics[width=\linewidth]{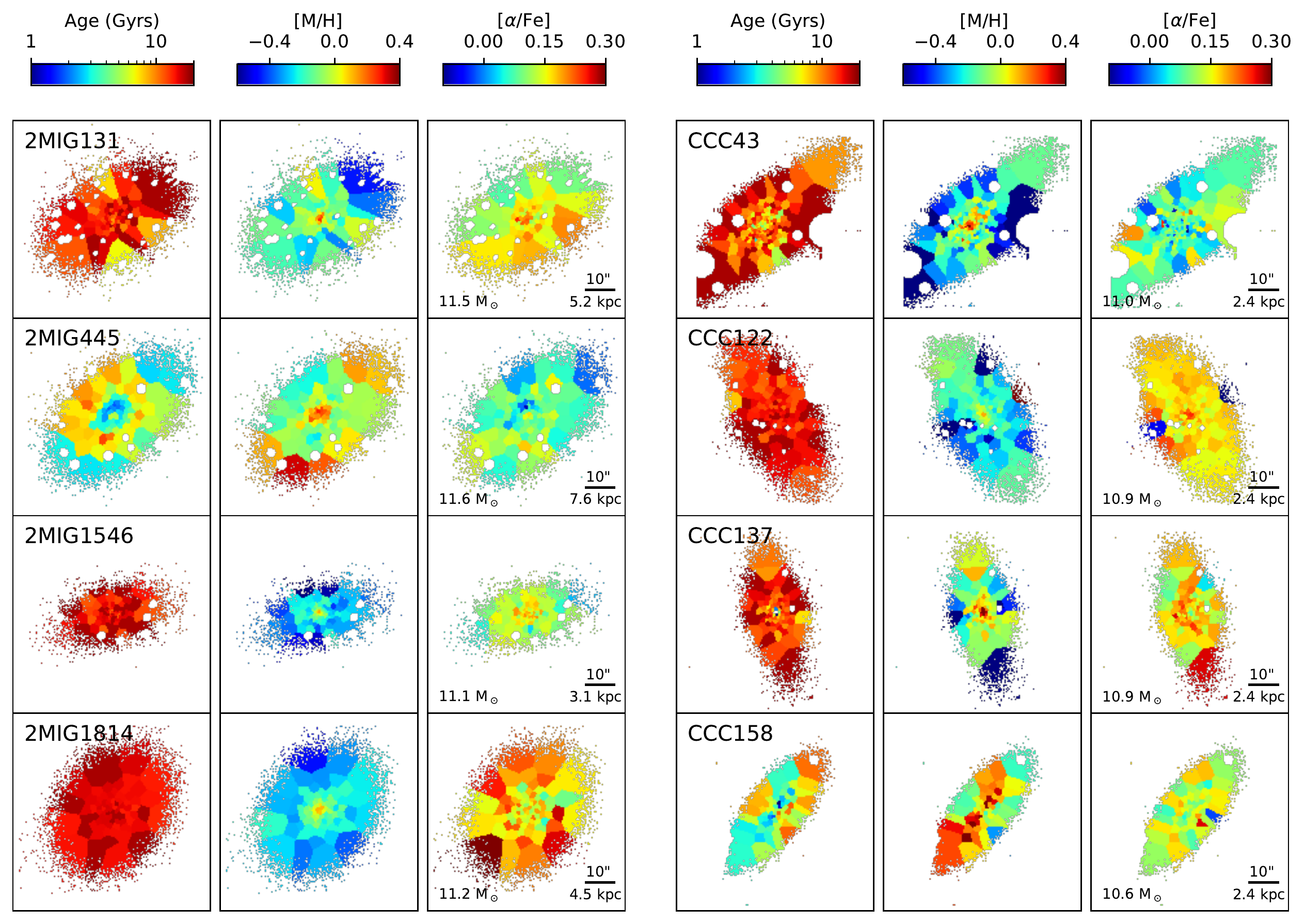}
 \caption{Age, metallicity and [$\alpha$/Fe] ratio maps for each galaxy, colour coded according to the colour bar at the top of each column. The isolated galaxies are in the left columns while the cluster members are in the right columns, and the label for each galaxy is in the age map. The scale bar at the bottom-right of the [$\alpha$/Fe] shows 10\arcsec\ and the corresponding physical size in kpc, and the $log_{10}$ of the mass is given in the bottom left of these maps. All maps are orientated with North up and East left and are plotted to the same scale.}
 \label{fig:radial_stellar_pops_maps}
\end{figure*}

\begin{figure*}
 \includegraphics[width=0.9\linewidth]{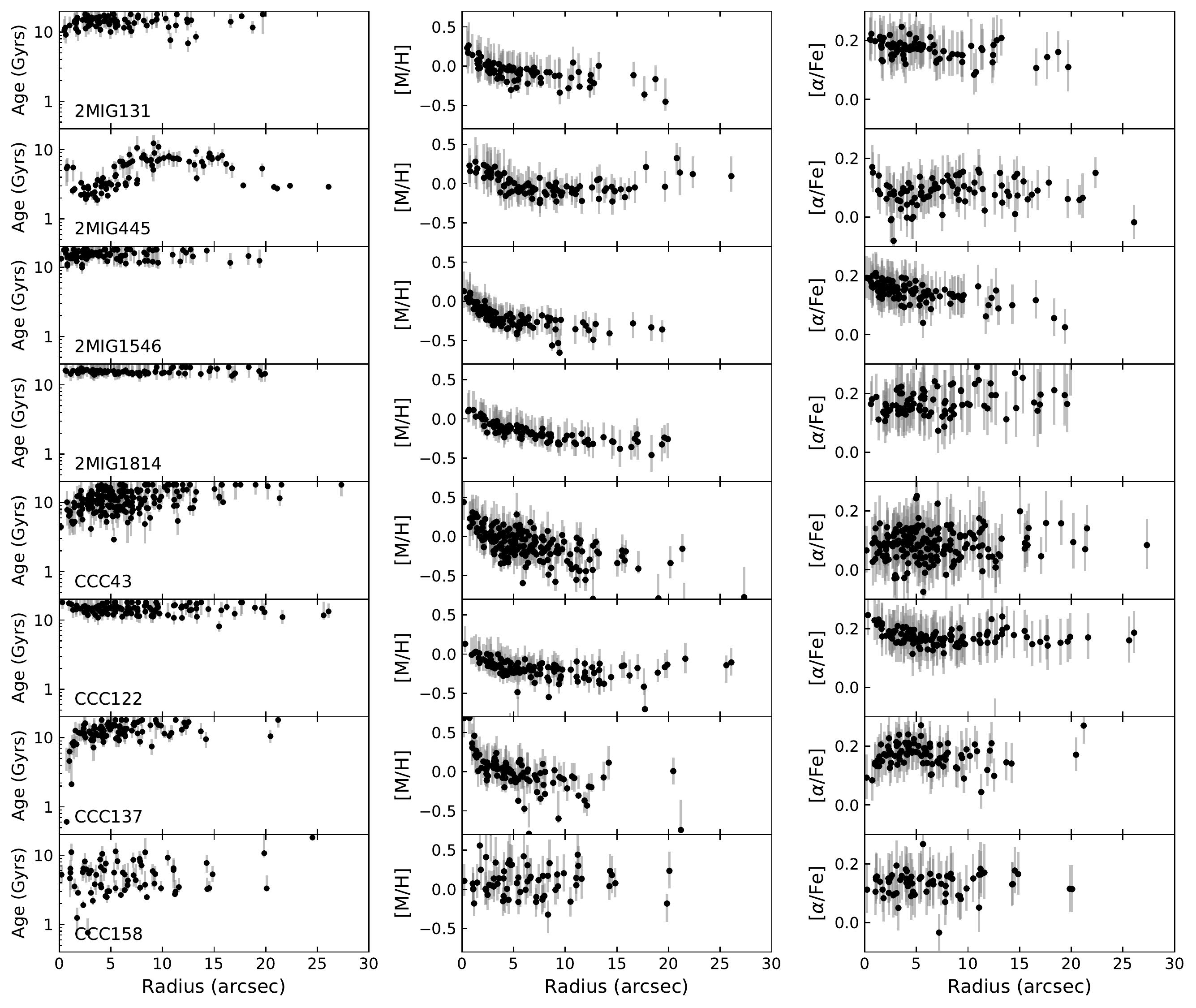}
 \caption{Age, metallicity and [$\alpha$/Fe] ratio plotted against radius for each galaxy, where the radius has been corrected for the inclination of the galaxy. The grey lines represent the uncertainties on each point.}
 \label{fig:radial_stellar_pops_plots}
\end{figure*}

\subsection{Luminosity-Weighted Ages and Metallicities}\label{sec:LW age met}

Since the light from a galaxy is dominated by the light emitted from the youngest stars, even if they contribute towards only a small fraction of the total galaxy mass, the luminosity-weighted stellar populations are often more of an indicator of how long ago the most recent episode of star formation was truncated. Thus, through the analysis of the luminosity-weighted stellar populations across each galaxy, one can determine where in the galaxy the star formation was truncated most recently, which in turn can provide clues as to the process which caused the gas to be stripped out and/or used up. 

The datacubes for each galaxy were first binned using the Voronoi Tesselation technique of \citet{Cappellari_2003} in order to extract spectra from across the galaxy with a higher Signal-to-Noise (S/N$>50$). The Lick indices for the hydrogen (H$\beta$), magnesium (Mgb) and iron (Fe5270 and Fe5335) were then measured in these spectra using the definitions of \citet{Worthey_1994} with the \textsc{indexf} software of \citet{Cardiel_2010}, and used as age (H$\beta$) and metallicity ($\text{[MgFe]$'$}=\sqrt{\text{Mg}b\ (0.72 \times \text{Fe}5270 + 0.28 \times \text{Fe}5335)}$) indicators. This software estimated uncertainties in these measurements through a series of simulations using the errors in the line-of-sight velocities and the S/N measured from the spectrum itself, details of which can be found in \citet{Cardiel_1998}. Where emission lines were detected in the spectra, the affected spectra were modelled again using \textsc{ppxf} to fit both the stellar absorption and gas emission spectra.  The best fit model for the emission lines was then used to correct for the emission in the original spectrum. 

The measurements were plotted onto the Single Stellar Populations (SSP) model grids of \citet{Vazdekis_2010} in order to convert them into estimates of the luminosity-weighted stellar populations. These models grids use the MILES stellar library \citep{Sanchez_2006} to create predictions of the line strengths for single stellar populations covering a wide range in age ($0.50-15.85$\,Gyr) and metallicity ($-2.32<[M/H]<0.22$), and were broadened to match the spectral resolution of the data using the MILES webtool\footnote{http://www.iac.es/proyecto/miles/pages/webtools/tune-ssp-models.php}. As a result, this step minimizes the loss of information that normally occurs when degrading the data to match lower-resolution models. 

The results for the luminosity-weighted ages and metallicities are plotted as maps in Fig.~\ref{fig:radial_stellar_pops_maps}, and as a function of radius along the major axis in Fig.~\ref{fig:radial_stellar_pops_plots} after correcting for the inclination of the galaxy. The error bars in Fig.~\ref{fig:radial_stellar_pops_plots} represent the effect on the line strengths due to uncertainties in the line-of-sight velocities and the S/N of each spectrum. The ages show no clear systematic trend, with most galaxies showing a flat age gradient and generally very old ages where small differences are hard to detect. The cores of many galaxies show hints of younger stellar populations in their cores (e.g. 2MIG~131, CCC~43 and CCC~137) or immediately outside their core (2MIG~445), which may indicate that the final episode of star formation within these galaxies occurred in their innermost regions. However, in most cases, even these younger stellar populations are above 10~Gyr, where age differences leave only very small effects on the line strengths, and so one must take care when interpreting these trends.

Two galaxies, 2MIG~445 and CCC~158,  show significantly younger stellar populations than the rest of the sample. While S0s are generally observed to contain older stellar populations, younger luminosity-weighted stellar populations have been detected in several studies \citep[e.g.][]{Kuntschner_1998, Poggianti_2001, Denicolo_2005}. The age plot for CCC~158 shows no strong gradient, but the map appears to show that the younger stellar populations lie along the major-axis in the inner regions of the galaxy, while the older stellar populations are seen in the outermost bins. 2MIG~445 on the other hand shows a ring of younger stellar populations circling a core of intermediate-age stars with an outer radius of $\sim12\arcsec$\ ($\sim9.1$\,kpc), with intermediate ages throughout the rest of the galaxy. These results indicate more recent star formation within these galaxies, potentially indicating more recent transformation from spirals, and the properties of these galaxies will be revisited throughout this paper. 

The metallicity maps and plots show evidence that all galaxies have higher metallicities in their cores than in the outskirts.  Similar metallicity trends have been seen before in studies of S0s,  e.g. \citep{Ogando_2005, Bedregal_2011, Prochaska_2011}.  \citet{Zibetti_2020}  suggested that the steeper negative metallicity trend in the inner regions of galaxies was created in the early dissipative collapse in the core of the galaxy, while the flatter gradients in the outer regions are an effect of the accretion of ex-situ stars from low-metallicity satellites. Interestingly, the maps in Fig.~\ref{fig:radial_stellar_pops_maps} show that in CCC~158 and 2MIG~445 the regions of higher metallicity coincide with the youngest stellar populations, indicating that the most recent episode of star formation was fuelled by gas that was likely enriched by star formation throughout the rest of the galaxy.

\subsection{Luminosity-Weighted $\alpha$-enhancement}\label{sec:LW alpha}

While the luminosity-weighted stellar populations presented in the previous section give details about the most recent episode of star formation within each component, the $\alpha$-element enhancement can provide information on the star-formation timescale of that event and the origin of the gas that fuelled it.  The $\alpha$-enhancement is measured using the ratios of $\alpha$-element-sensitive indices, in this case Mgb, to $\langle \text{Fe} \rangle=(\text{Fe}_{5270}+\text{Fe}_{5335})/2$. $\alpha$ elements are released into the interstellar medium by Type~II supernovae, which start occurring shortly after the onset of star formation, while Fe originates in Type~Ia supernovae, which only start $\sim1$Gyr after the onset of star formation. Consequently, the ratio of these elements can be used as an indicator of the star-formation timescale, such that shorter episodes of star formation will give an $\alpha$-enhanced stellar population due to the enrichment of magnesium from the SNII, and the $\alpha$-enhancement will begin to drop after SNIa appear due to the dilution of magnesium with iron in the interstellar medium. For example, it has been found that the highest [$\alpha$/Fe] ratios are achieved in galaxies with the shortest  half-mass formation time \citep[<2~Gyr;][]{delaRosa_2011}, and that the [$\alpha$/Fe] decreases with increasing star-formation timescale out to $\sim10\,$Gyr \citep{Wiersma_2009}.  

To measure the $\alpha$-enhancement, the Mgb and $\langle \text{Fe} \rangle$ line strengths measured from the binned spectra were plotted onto the SSP model grids of \citet[][hereafter TMJ models]{Thomas_2011}. These models give estimates of the relative metallicity and [$\alpha$/Fe] ratio for a range of ages of stellar populations, and for each galaxy the model was selected that most closely matched the mean luminosity-weighted age for galaxy. Since the \citetalias{Thomas_2011} models are based on the MILES stellar library, the spectral resolution is similar to that of the MUSE data, and so no corrections were necessary to account for the difference in resolution. However, with velocity dispersions of between $130-300$~\kms\ for each galaxy, the line strengths needed to be corrected for this broadening. The correction was carried out by fitting each spectrum with \textsc{ppxf} and using the weights for each template spectrum used in the fit to obtain a best fit spectrum at the resolution of the MILES library, i.e. without including for the velocity dispersion of the data. This model spectrum was then convolved with a series of Gaussians to simulate the broadening for a range of velocity dispersions, and the strengths of each line measured in the same way from these broadened spectra. Finally, the necessary correction was calculated for the measured velocity dispersion from the galaxy spectrum via interpolation. The corrected line strengths were then plotted onto the grids to obtain estimates of the $\alpha$-enhancement, and the results are given in the right columns of Fig.~\ref{fig:radial_stellar_pops_maps} and \ref{fig:radial_stellar_pops_plots}.

It can be seen that the [$\alpha$/Fe] ratios are relatively flat with radius across most galaxies, with 2MIG~131 and 2MIG~1546 showing a weak negative gradient with radius. Additionally,  variations in the [$\alpha$/Fe] ratios are detected in the cores of several galaxies, with an increase in the inner $2\arcsec$\ ($0.5$\,kpc) of CCC~122 that coincides with the older, more metal-rich stellar populations, and a drop in the inner $2\arcsec$\ ($0.5$\,kpc) of CCC~137 where the significantly younger, more metal-rich populations are found. These results hint again at different star-formation histories in the cores of these galaxies compared to their outer regions.

The [$\alpha$/Fe] ratio for 2MIG~445 is particularly interesting, with a similar shape to the age plot and map within $5\arcsec$. Together, these results indicate that the light in the very core of this galaxy is dominated by stars created during a short episode of star formation, surrounded by a ring of younger stars that were formed during a more extended period of star formation.

However, the strength of all of these features and their uncertainties are of a similar magnitude to the scatter in the measurements, and little can be said for certain without looking at the properties in each  component within the galaxies.


\section{Extracting bulge and disc spectra with BUDDI}\label{sec:buddi}

Clues to how S0s have transformed from their progenitor galaxies and which processes led to the truncation of star formation lie in the stellar populations of their bulges and discs. However, in the inner regions of S0 and spiral galaxies, the light from the disc is always superimposed upon the bulge, and so the radial stellar populations represent the light from both components in varying fractions \citep[e.g.][]{Kennedy_2016b}. As a result, cleanly extracting the spectra from either of these components is difficult, and if an incorrect model is used, the stellar populations analysis can suffer from contamination of light from other components.

One way to cleanly extract the spectra of individual components within galaxies is to create a wavelength-dependent model of the light profile of each component with \textsc{buddi} \citep[Bulge--Disc decomposition of IFU data;][]{Johnston_2017}. \textsc{buddi} uses a modified form of \textsc{Galfit} \citep{Peng_2002, Peng_2010} called \textsc{GalfitM} \citep{Haeussler_2013, Vika_2013} to model the light profile of multi-waveband images of a galaxy simultaneously. User-defined Chebychev polynomials are implemented to account for variations in the structural parameters of each component as a function of wavelength, thus using information from all the images for each fit. As a result, the S/N of the data set is boosted over that of any individual image, thus allowing reasonable estimates for the structural parameters to be made for images with lower S/N. While this technique was initially developed for large scale surveys with a handful of wavelengths/bands, it is particularly powerful for IFU datacubes, which can be considered as a series of narrow-band images of the target at each wavelength step in the spectral direction, as it can produce reliable fits to image slices with low S/N or which are affected by sky lines. While \textsc{buddi} was originally written to model the bulges and discs of galaxies, it has been successfully applied to fits to extreme galaxy components, from extended stellar haloes \citep{Johnston_2018} to the nuclear star clusters within dwarf galaxies \citep{Johnston_2020}. 

It is important to note that \textsc{buddi} uses only the light profile information of the galaxy within the datacube to create the wavelength-dependent models for each component and thus extract their clean spectra. No information is provided or extracted by \textsc{buddi}  regarding the spectra or the line strengths. Consequently, the spectra extracted for each component is independent of any previous assumptions or measurements of the stellar populations, flux fractions or mass fractions of each component. The main bias that can be introduced by the fits is the number of components and their starting parameters, and to combat this issue, users are encouraged to repeat steps~2 and 3 below to ensure the best fit has been achieved. 

A full description of \textsc{buddi} can be found in \citet{Johnston_2017}, with a particularly useful flow chart in Fig.~1 of that paper, and a brief overview is presented below.

\subsection{Step 0: Create the mask and PSF profile}\label{sec:step_0}
Before modelling a galaxy with \textsc{buddi}, a few preparations are needed for the data, such as creating a bad pixel mask to mask out pixels within the image that fall outside of the MUSE field-of-view (FOV). Some of the datacubes also contain light from other sources, such as foreground stars or background galaxies. Generally it is preferable to fit these objects simultaneously to better model for the light from the target galaxy. However, in this study it was decided to mask out these objects since including models for even just the brightest objects within the FOV would have significantly slowed down the fits. Furthermore, since the kinematics corrections described in the next step are only applied to the dominant kinematical component, any distortion in the kinematics from these objects would likely lead to contamination of the spectra extracted from the target galaxy.

In order to achieve a good fit to the image of a galaxy, the model must be convolved with a point-spread function (PSF) profile that matches the image quality of the data. Consequently, to achieve a good fit with \textsc{buddi}, a PSF datacube must be provided.  Ideally this model datacube would be created from non-saturated stars within the FOV of the science image, but the large size of the majority of the galaxies within the MUSE FOV and the lack of isolated stars in the fields meant that this option was not always possible. In these cases, between 1 and 6 stars were identified in the corresponding sky exposures, which were observed as part of the same observing block as the galaxy exposures and thus display similar image quality to the science data. In rare cases where the sky exposure contained no suitable stars, the standard star field was used. In such cases, the standard star was found to be very bright, with excess light spreading horizontally in the images within the slicers, and so other, fainter stars within the FOV were used to create the PSF datacube, generally between 1 and 4 stars. Since the standard stars were observed at different times during the night and in different parts of the sky, a single standard star exposure was selected for each target datacube that best matched the image quality of the corresponding data, as measured from the FWHM of a point source in the white light image of the target datacube. These model PSF profiles were tested by modelling a star or other point source within the FOV of the science datacube and evaluating the residual images. While this approximation of the PSF is not perfect, it was found to provide a satisfactory fit for each datacube. 

\subsection{Step 1: Obliterate the kinematics}\label{sec:step_1}
\textsc{GalfitM} can only fit symmetric models to an image of a galaxy, and with the narrow wavebands of image slices in an IFU datacube, one must be careful when modelling image slices at wavelengths of strong spectral features. For example, the rotation of a galaxy can mean that part of the image slice might reflect the flux of the continuum while another part may be in an absorption feature, resulting in an asymmetric image. Additionally, the velocity dispersion of a galaxy may decrease with radius, meaning that spaxels from the centre of the galaxy are from within the wings of a spectral feature while those at larger radii are within the continuum. Both these effects can seriously affect the fits to the affected image slices, resulting in messy spectra. Therefore, the first step carried out by \textsc{buddi} is to measure and smooth out the kinematics across the galaxy in order to reduce these effects. 

For this step, the datacubes were binned using the Voronoi tessellation technique of \citet{Cappellari_2003}, and the kinematics of the binned spectra were measured using the penalised Pixel Fitting software (pPXF) of \citet{Cappellari_2004} using the template spectra of \citet{Vazdekis_2010}, which are based on the MILES stellar library of \citet{Sanchez_2006} and consist of 156 template spectra ranging in metallicity from $-1.71$ to $+0.22$, and in age from 1 to 17.78 Gyr. The spectra in each spaxel within the datacube were then shifted to correct for the line-of-sight velocity, and broadened to match the maximum velocity dispersion measured within the galaxy. The kinematics across this sample of galaxies are discussed in detail in \citetalias{Coccato_2020}.

\subsection{Step 2: Fit the white-light image}\label{sec:step_2}
The first step in the actual light-profile fitting process was to model the white-light image of the galaxy with \textsc{GalfitM}. This step allows the user to quickly assess the fit in terms of the number of components necessary and the starting parameters before moving onto the more time-consuming steps to fit the images throughout the rest of the datacube. Each fit started with a uniform sky background, hereafter referred to as the sky component, and a single S\'ersic profile, with additional S\'ersic profiles added successively to the model until the best fit was achieved. The best fit was decided by looking at the residual images, which were created by subtracting the best fit model from the input image, to find the balance between including enough model components for a good fit while at the same time using the minimum number of components necessary to create a good fit.

\begin{figure*}
 \includegraphics[width=0.9\linewidth]{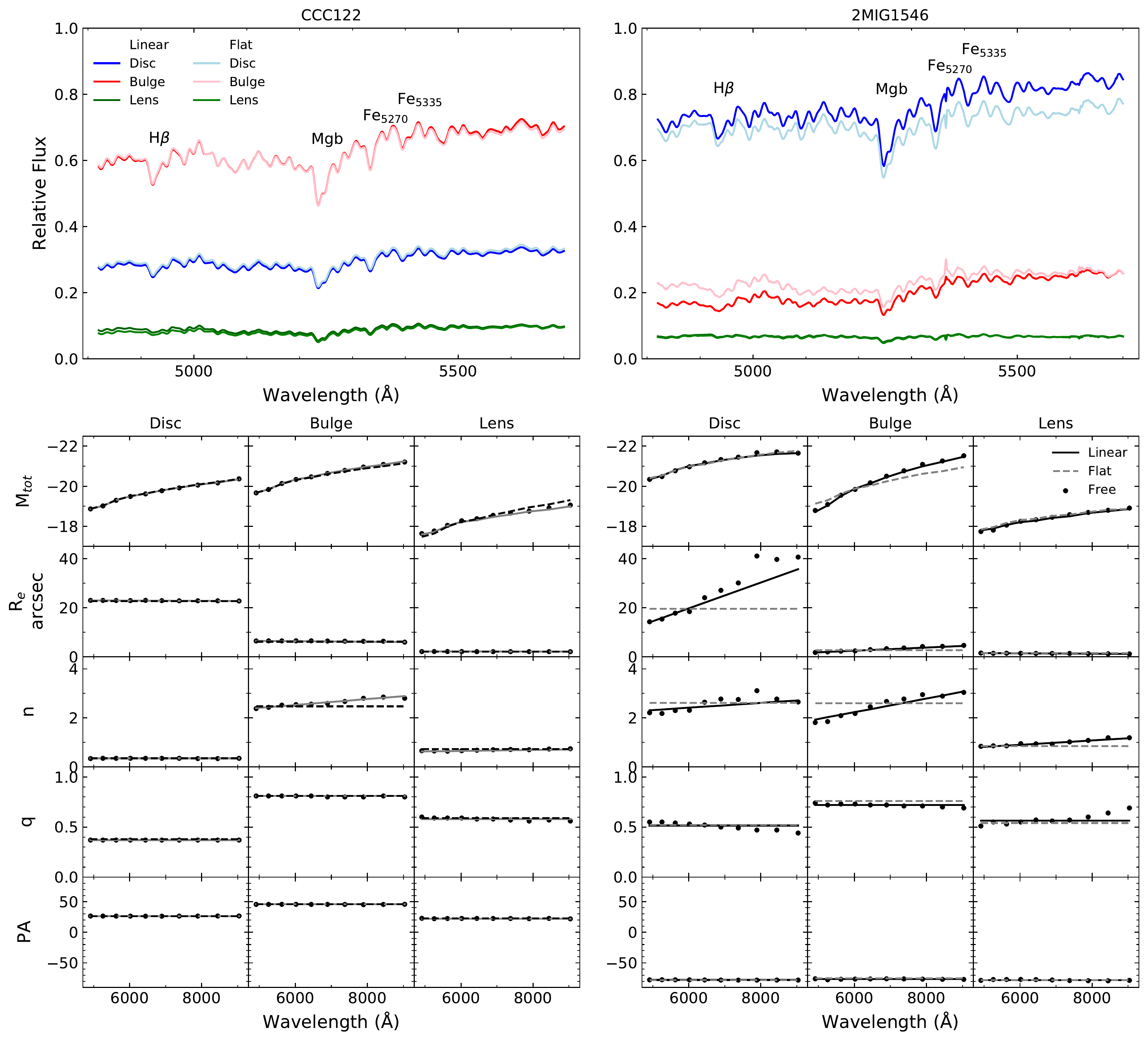}
 \caption{A comparison of the fit parameters for free, flat and linear fits for CCC~122 on the left and 2MIG~1546 on the right. The top plots show the differences in the spectra for each component between the flat and linear fits (lighter and darker lines respectively).  The bottom plots show a comparison of the fit parameters when they are allowed complete freedom (points), held fixed with wavelength (dashed line) and allowed to vary linearly with wavelength (solid line). The black and grey lined identify the fits that were ultimately selected and rejected. the magnitudes represent the absolute integrated AB magnitudes. }
 \label{fig:comp_flat_lin}
\end{figure*}

\subsection{Step 3: Fit the narrow-band images}\label{sec:step_3}
The datacube was then rebinned into a series of 10 narrow-band images with high S/N spanning the full spectral range of the MUSE datacube, and these images were modelled with \textsc{GalfitM} using the best fit to the white light image as the starting parameters. The first fit was completely unconstrained, simulating the results obtained by running \textsc{Galfit} on each narrow-band image independently and giving an idea of the variations in the parameters as a function of wavelength. The fit was then repeated using Chebychev polynomials to constrain the parameters for effective radius ($R_e$), S\'ersic index ($n$), axis ratio ($q$) and position angle ($PA$) as a function of wavelength while leaving the integrated magnitude ($m_{tot}$) free. This step is repeated if necessary until the Chebychev polynomials give a good representation of how each parameter varies with wavelength and the residual images show a good fit at all wavelengths. This step defines the polynomials for all parameters but the magnitudes in the next step of the fitting process. In general, the $PA$ and $q$ were modelled using polynomials of order 1 (flat with wavelength) for all galaxies, and $R_e$ and $n$ were modelled with both polynomials of order 1 and 2 (linear variation with wavelength). Examples of these polynomial fits to the parameters from the free fits are given in Fig.~\ref{fig:comp_flat_lin} for CCC~122 and 2MIG~1546, showing cases where the fits using orders 1 and 2 result in small and large differences in the final spectra. For example, on can see that the two fits to 2MIG~1546 result in up to a $\sim10\%$ difference in the light fractions and SEDs for the bulge and disc components in the blue end of the spectrum, while these differences are almost negligible in CCC~122. In 6 of the eight galaxies, the fits using these three different orders converged on similar fits, while in 2MIG~131 and 2MIG~1814 the fits using a polynomial of order 1 resulted in very different structural parameters

\begin{figure*}
 \includegraphics[width=\linewidth]{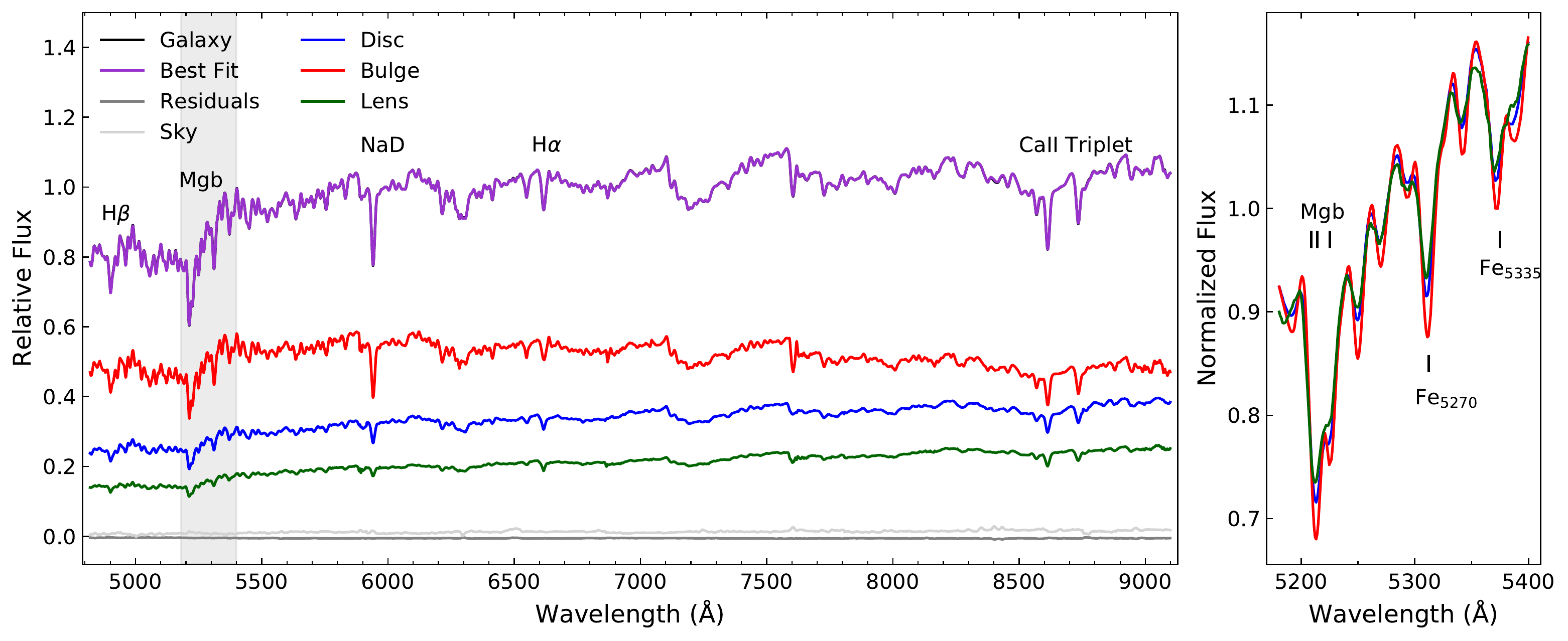}
 \caption{\textit{Left}: the spectra extracted for the disc (blue), bulge (red) and lens (green) components in CCC~137, plotted relative to the total flux of the galaxy within the FOV of the datacube. The purple line represents the combined bulge+disc+lens+sky spectrum, superimposed upon the integrated spectrum of the galaxy in black, and the dark and light grey lines represent the residuals from faint asymmetric features and background objects and the sky background respectively. \textit{Right}: a zoom-in of the normalised bulge, disc and lens spectra over the spectral region marked by the grey band on the left plot, showing the differences in the line strengths over the Mgb and Fe absorption lines.}
 \label{fig:spectra}
\end{figure*}

In the case of 2MIG~131, the fits using a polynomial order 1 resulted in a more compact bulge with a lower S\'ersic index, and the light fractions for the spectra from each component varied by up to $\sim40\%$. On the other hand, the flat fit to 2MIG~1814 converged on a model with a very extended, very faint component which was $\sim4-5$ magnitudes fainter than the disc component in any other fit. Tests were carried out on these two galaxies, modelling their narrow-band images with different combinations of initial parameters and polynomial fits, and it was found that these different fits were derived only when all the parameters except the magnitudes were modelled with polynomials of order 1. Consequently, it was decided that this order of polynomial induced too much restriction in the fit, and that the final fits were untrustworthy. Consequently, for these galaxies, the flat fits were carried out with polynomials of order 2 for $R_e$ and order 1 for $n$, $q$ and $PA$.

Ideally, at this step, one wants to select the lowest order of polynomial that provides the best match to the results from the free fit. In Fig.~\ref{fig:comp_flat_lin}, an order of 1 appears to provide a good fit to CCC~122, while 2MIG~1546 clearly requires a higher order for the disc $R_e$ and bulge and disc $n$. However, in this study, the small number of galaxies meant that the fits to the image slices described in the next step could be carried out using both orders, thus allowing a comparison of the spectra and stellar populations for each component before deciding which order to use for the analysis.

\subsection{Step 4: Fit the image slices and extract the decomposed spectra}\label{sec:step_4}
The final step of the process is to find the best fit model for every image slice within the datacube, and to use the results from these fits to derive the clean spectra for each component.  To reduce the required computing time for this step, the datacube was split up into batches of 10 consecutive image slices, which were modelled by \textsc{GalfitM} simultaneously, with the parameters for the $R_e$, $n$, $q$\ and $PA$\ held fixed as a function of wavelength according to polynomials derived in the previous step. This constraint on the model ensures a reliable fit to the galaxy in each image slice since the S/N of the individual images are lower than that of the stacked images used in the previous two steps, while also preventing artefacts appearing in the fit parameters at wavelengths where each batch of 10 image slices meet. The final parameter included in the fit is the integrated magnitude of each component, which is allowed to vary with complete freedom in each image slice. Consequently, the integrated magnitudes derived reflect total flux of each component at the wavelengths of the image slices used in that fit. By plotting these values for the total flux as a function of wavelength for each component, their clean 1-dimensional spectra are created, which are integrated to infinity. As described above, \textsc{buddi} uses no information on the spectral properties as part of the fit, simply seeing a series of images to model. As a result, the spectra created by \textsc{buddi} are free from bias on the stellar populations or line strengths within the galaxy. Additionally, the combination of the 2-dimensional fit and the polynomials used for the structural parameters lead to consistent fits to each image slice, and thus very little noise in the final spectrum, especially for large, bright galaxies like those used in this study.

An example of these spectra extracted for each component in CCC~137 is given in Fig.~\ref{fig:spectra}, showing the  fluxes of each component relative to the total flux from the galaxy. These final spectra represent the clean model spectra for all the components included in the fit, and can be used for stellar populations analysis of those structures. This figure also plots the sky spectrum, which was created from the sky component included in the fit, and the residual  spectrum, which reflects the remaining flux after subtracting the combined best-fit spectrum from the total flux of the galaxy.  The residual spectrum consists of light from faint asymmetric features within the galaxy and background objects within the FOV, which will be discussed in more detail in Section~\ref{sec:residuals}.

Figure~\ref{fig:spectra} also shows a zoom-in of the spectral region covering the Mgb and Fe lines used for the luminosity-weighted stellar populations analysis in Section~\ref{sec:LW_pops}, showing the normalised spectra for the bulge, disc and lens of CCC~137 superimposed. It can be seen that the bulge spectrum shows stronger lines, indicating a higher metallicity, followed by the disc and then the lens. Additionally, a difference in the shape of the Mgb~triplet can be seen between the bulge and lens spectra due to the different strengths of the Mg$_{5183.6}$ line to the right of the triplet.

By this stage, two spectra had been extracted for each component within each galaxy, as derived using the fits with polynomials of order 1 and 2. The analysis described in the following sections was carried out with both sets of spectra, and while small variations were found in the individual measurements, the same trends were seen with both data sets. In the end, the final decision on which set of spectra to use for each galaxy was based on a combination of the better fit to the structural parameters from the free fit (see Fig.~\ref{fig:comp_flat_lin}), fewer artefacts in the residual images (e.g. see Section~\ref{sec:residuals}), and the stellar populations. For example, where the line strengths fell off the Single Stellar Population (SSP) model grids used in Section~\ref{sec:LW_pops}, the order that led to the smallest offset between the line strengths in the spectrum and the nearest point in the grid was selected as the better fit. If these steps failed to identify the better fit, then the simpler model was used for the analysis. In this way, the spectra derived using the flat fits (order 1) for 2MIG~131, 2MIG445, CCC~43 and CCC~122, and the linear fits (order 2) for 2MIG~1546, 2MIG~1814, CCC~137 and CCC~158 were used for the analysis presented in the rest of this paper.


\section{Overview of the fits}\label{sec:overview_fits}
\subsection{The number and nature of the components in each galaxy}\label{sec:no_components}

The models used for each galaxy were created with no prior information from fits to photometric data in the literature to reduce bias during the fitting process. Unlike many other studies where galaxy light profiles are modelled \citep[e.g.][]{Simard_2011, Mendel_2014, Vika_2014, Bottrell_2019}, the fits were not restricted to single or double S\'ersic profiles, where the latter is taken to represent a bulge+disc fit.  Instead, the fits were started using a single S\'ersic profile, and the complexity of the model was built up by adding additional S\'ersic or PSF components until the best fit is obtained. The best fit was determined first by comparing the $\chi^2$, and then by studying the residual images of each fit to identify the model that gave a good fit to the data with the minimum number of components. For example, one can compare the flux levels in particularly bright or dark regions of the residual image, which represent areas where the light was under and over subtracted by the light of the model, or look at the standard deviation of the all the unmasked flux values in the residual image. Additionally, broad rings of alternating light and dark regions can be a characteristic signature of several issues, such as poor initial estimates for the structural parameters, that the light profile of the galaxy is cored and that a core-S\'ersic profile is required instead of a standard S\'ersic profile \citep[e.g.][]{Dullo_2014}, or that an additional component that is required for the fit. Finally, when the inclusion of an additional component showed negligible differences or improvements the residual image, it was considered unnecessary and the best fit was accepted as the fit with one less component.   It was found that all the galaxies in this sample achieved a better fit using at least three components, with an additional fourth component in the form of a PSF profile in the cases of 2MIG~1814, CCC~122 and CCC~137. The fits to all the galaxies are given in Appendix~\ref{sec:Appendix_A}, with the residual images from the 2-component fits given in Appendix~\ref{sec:Appendix_B} for comparison.
 It can be seen that in the 2-component fit, the fits to the light profile along the major axes are generally quite good, but the residual images reveal more artefacts, such as the rings of alternating light and dark regions outlined above. Thus, these images emphasise the importance of using the two-dimensional spatial information wherever possible. Similar findings were presented by \citet{Laurikainen_2009} for their fits to S0 galaxies, where better results were obtained when fitting 2-dimensional images over a one-dimensional light profile in cases where the galaxy contains more than simply a bulge and disc.

All galaxies were found to contain a clear extended disc structure alongside two more compact components with S\'ersic light distributions. Similarly complex models for the bulges have been seen before -- composite bulges, which are made up of  two or more structures such as pseudobulges, central discs or boxy/peanut bulges,  have been detected in small samples of galaxies by \citet{Erwin_2003}, \citet{Nowak_2010}, \citet{deLorenzo_2012} and \citet{Erwin_2015}, and a study by \citet{Mendez_2014} found that $70\%$ of their sample of barred disc galaxies host composite bulges.  The images of each component in Appendix~\ref{sec:Appendix_A} demonstrate the diversity of these inner components, with some showing rounder profiles while others are more elongated, and with a range of S\'ersic indices between 0.5 and 4.5. These structures likely represent a mix of classical and pseudobulges, as well as structures such as lenses, bars and thin discs. 

Lenses have been detected in photometric studies of S0s by  \citet{Laurikainen_2005, Laurikainen_2009}, and are distinguished from bulges by their flatter surface brightness profile and sharper edges. They have been found to have generally lower surface brightnesses than pseudobulges \citep{Kormendy_2004}. Bars and discs on the other hand can be identified through their kinematic signatures. For example, \citet{Krajnovic_2011} found that the distribution of $h_3$ against $v/\sigma$ can be used to distinguish between rapidly rotating galaxies with and without a bar or ring present in an IFU datacube. Furthermore, \citet{Guerou_2016} demonstrated that a clear trend in this distribution appears when the spaxels or bins corresponding to a nuclear disc are highlighted.

Having identified the disc in each galaxy, two more compact S\'ersic components remained. Within these components, the bulges were selected by eye as the more extended component of the two with a steeper surface brightness profile and the softer edge (i.e. less sharp drop in the flux in the outskirts), and in 6 of the eight galaxies this component was found to be the brightest structure at the core of the galaxy, with the exceptions being CCC~43 and CCC~158. The remaining component was found to be elongated in around half of the galaxies-- 2MIG~445, 2MIG~1546, CCC~122 and CC158-- and relatively circular in the remaining galaxies. This component may represent either  a bar, an inner embedded disk or a lens. None of the galaxies show evidence of a strong bar in their visual appearance, and no kinematical signatures were detected in \citetalias{Coccato_2020}. In the case of an inner disc, one would expect it to have a similar inclination and position angle as the outer disc, and would expect to find a kinematical signature of a rotating disc within the dispersion-supported bulge. Simulations by \citet{Eliche_2018} of S0 galaxies containing lenses viewed at different inclinations found that the lenses can show a range of shapes when viewed edge-on, with the majority appearing vertically thin and disc like ($\sim64\%$) while the rest displayed vertically thick lentil and spheroidal shapes. Consequently, the inclination angles were not used in this study, and instead the kinematics were used to distinguish between a lens and an inner disc.

The $h_3$ Gauss-Hermite polynomial parameter gives a measure of the skewness in the line-of-sight velocity distribution (LOSVD), and is generally found to anti-correlate with the velocity curve in rotating disc galaxies. If another kinematical component is present, its distinct velocity distribution will further affect the shape of the LOSVD of the galaxy as a whole in that region. Therefore, by plotting $h_3$ against the $v/\sigma$, one can look for evidence of additional kinematic components.

\begin{figure}
 \includegraphics[width=\linewidth]{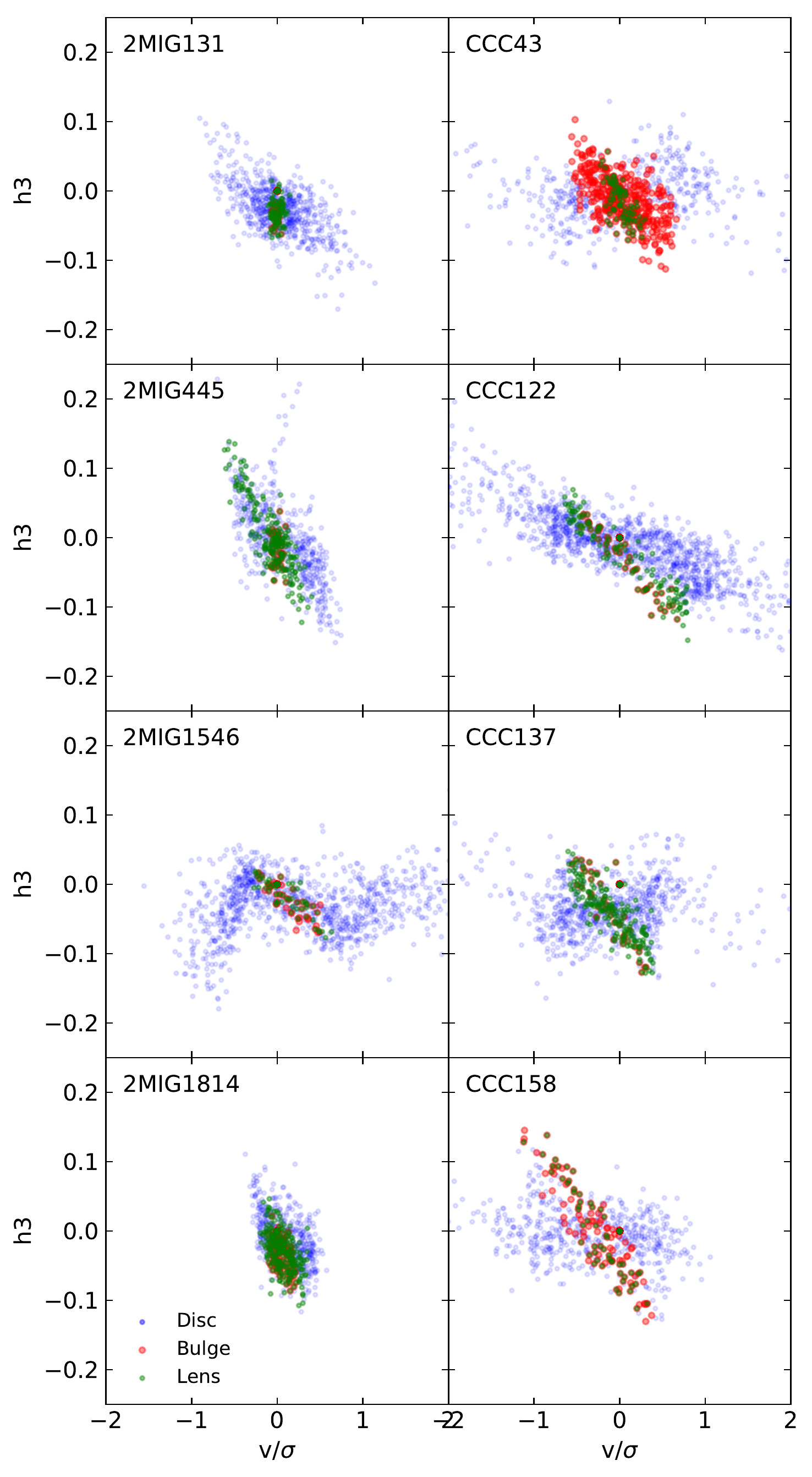}
 \caption{The plots for $v/\sigma$ versus $h_3$ for each binned spectrum, with those within the disc, bulge and lens components highlighted in blue, red and green respectively. The isolated galaxies are on the left while the cluster galaxies are on the right.}
 \label{fig:v_sig}
\end{figure}

Using the kinematics information from \citetalias{Coccato_2020}, the plots for $h_3$ against $v/\sigma$ are given in Fig.~\ref{fig:v_sig} for each galaxy, where the measurements from within the disc, bulge and lens/inner disc regions have been highlighted in blue, red and green respectively.  For clarity, the measurements marked as bulges and lenses indicate only those binned spectra where the mean flux of each spaxel in that bin are greater than half of the peak flux of that component in the white-light image of the galaxy. In general, all galaxies show an anti-correlation between $h_3$ and $v/\sigma$, as expected, but the red and green points appear to show distinct trends within the plots for all the cluster galaxies that are similar to that seen for the nuclear disc in NGC~3115 by  \citet{Guerou_2016}. 
These trends indicate that the kinematics are dominated by a fast rotating disc, but have a significant tail towards lower velocities due to the presence of the slower rotating bulge in the inner regions of the galaxy. If the kinematic signature of an inner disc was present, one would expect to see distinct trends in this figure between the bulge and this disc due to the bulge being larger and higher above the plane of the disc. Since no clear differences can be seen in the kinematics of the bulges and lenses/inner discs in these plots, we can only say that if an inner disc is present, its kinematic signatures  are masked by those of the bulge and outer disc. The fact that these trends are only seen in the cluster galaxies  may reflect a dependence on environment, but could also be a result of these galaxies being more rotationally supported, making such trends easier to see.

\begin{figure*}
 \includegraphics[width=0.8\linewidth]{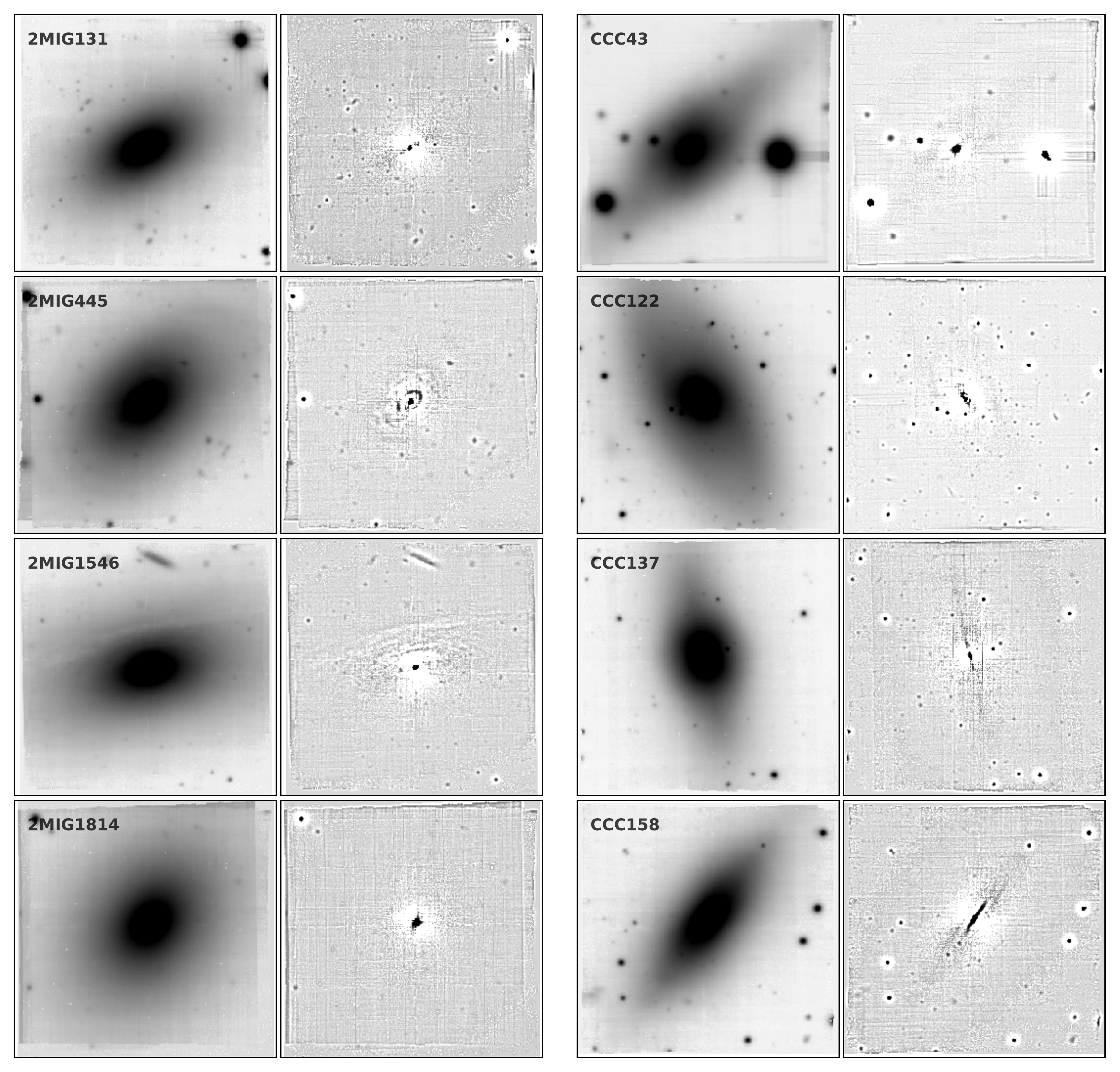}
 \caption{White-light images of each galaxy alongside the unsharp masked images. The galaxies on the left are the isolated galaxies while those on the right are the cluster galaxies. }
 \label{fig:unsharp_mask}
\end{figure*}

At this point it is interesting to note the unusual relationship between $h_3$ and $v/\sigma$ in 2MIG~1546, which shows a correlation at larger radii and an anti-correlation at lower radii. This trend indicates that in the outskirts of the galaxy an additional kinematic component is present that is rotating faster than the main disc component but never dominates the light in those regions. Such a phenomenon would result in a tail towards higher velocities in the LOSVD. In the inner regions, on the other hand, the tail in the LOSVD switches to lower velocities, which reflects the contribution of the slower-rotating bulge kinematics to the LOSVD.  The velocity and velocity dispersion maps for this galaxy in \citetalias{Coccato_2020} appear relatively smooth, though a distortion in the $h_3$ and $h_4$ parameters can be seen to the west of the centre of the galaxy along the major axis. While not discussed in that paper, this feature could reflect this additional kinematic component, which may be an artefact of a minor merger in the history of the galaxy

Another technique used by \citet{Laurikainen_2005} to distinguish between lenses and inner discs is to apply unsharp masking to an image of the target, which can reveal structures such as spiral arms and edge-on discs. This technique was thus applied to the white-light images of each galaxy using a Gaussian with a kernel of 4 pixels. These images are given in Fig.~\ref{fig:unsharp_mask}, and in CCC~158 it is possible to see evidence of an edge-on disc in the core of the galaxy. Further evidence for a nuclear disc would be a drop in the velocity dispersion and an anticorrelation in the $h_3$ moment within that region. The kinematics maps for each galaxy are presented in the Appendix of \citetalias{Coccato_2020}, and it is possible to see both these signatures in the maps for CCC~158. These kinematic signatures correspond approximately with the light profile of the third component in this galaxy, which is also brighter than the bulge in the core of the galaxy, thus indicating that the trend seen in Fig.~\ref{fig:v_sig} may instead reflect the kinematics of this component as opposed to the bulge. If this component is an inner disc, it may contain distinct stellar populations relative to the bulge and outer disc. Revisiting the stellar population maps in Fig.~\ref{fig:radial_stellar_pops_maps} show  particularly young and metal-rich stellar populations in this same region, though in this case the ellipse of younger stellar populations has a length of $\sim15\arcsec$\ or 3.6~kpc while the effective radius of the third component is $\sim10\arcsec$. Since younger stars tend to dominate the light from a galaxy, it is possible that this third component appears more extended in the stellar population maps because it contains younger stars than the other two components.  Consequently, it is possible that this component in CCC~158 may be an embedded inner  disc seen at high inclination, while in all the other galaxies in the sample this component is likely to be a lens. However, for the rest of this paper, we will refer to this component in all galaxies as a lens unless discussing CCC~158 explicitly.

\subsection{Structural parameters of each component}\label{sec:parameters}

Having identified the discs, bulges and lenses, it is now possible to look at their structural parameters in more detail.  In recent years, bulges have been separated into two categories: classical and pseudobulges, where classical bulges are typically dispersion-supported spheroids with higher S\'ersic indices while pseudobulges are rotationally supported, discy and with lower S\'ersic indices. \citet{Kormendy_2004} use the information for the bulge vs disc ellipticity and the kinematics of the galaxy in the form $v/\sigma$ to distinguish between these classifications, while \citet{Fisher_2008} define a S\'ersic index of $n=2$ to separate the morphologies, and \citet{Fabricius_2012} use a combination of the kinematics and the properties derived through bulge--disc decomposition. Since no standard system is common throughout the literature, we have chosen to adopt the  \citet{Fisher_2008} technique in this work.

\begin{figure}
 \includegraphics[width=\columnwidth]{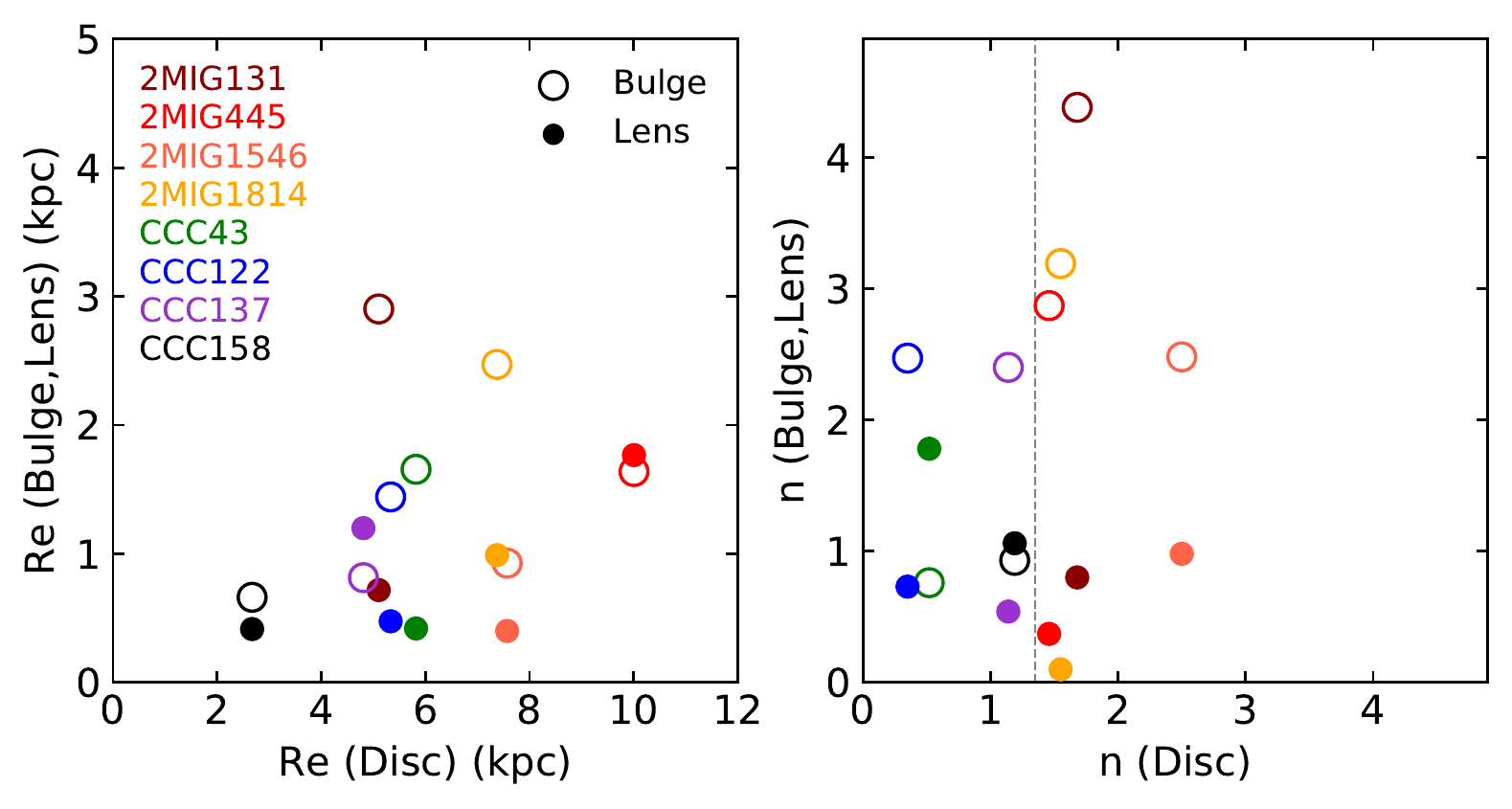}
 \caption{A comparison of the effective radii and S\'ersic indices of the bulges and lenses versus the discs. The comparisons with the bulges are represented by the hollow circles while the lenses are filled circles. }
 \label{fig:Re_n}
\end{figure}

Figure~\ref{fig:Re_n} plots the measurements for the $R_e$ and $n$ for each bulge and lens component against that of the corresponding disc. Where the $R_e$ was modelled with a Chebychev polynomial of order 2, the value of $R_e$ was taken from the fit to the narrow-band image closest to the centre of the MUSE spectral range. It can be seen that 6 of the galaxies have bulges with $n>2.0$, classifying them as classical bulges while the other two, CCC~43 and CCC~158, are pseudobulges. It is interesting to note that these two galaxies are the only cases in which the lenses dominate the light at the core of the galaxy, which is inconsistent with the findings of \citet{Kormendy_2004} that lenses tend to have lower surface brightnesses than pseudobulges. Using  the kinematics information derived in \citetalias{Coccato_2020}, it was also found that the central velocity dispersion in each galaxy correlated with the bulge morphology, such that the classical bulges have $\sigma_0>200$\kms\ while the psuedobulges have $\sigma_0<200$\kms. 

Furthermore, all the lenses and inner discs have $n<2.0$, and with the exception of CCC~43 and CCC~158, the bulges all have higher S\'ersic indices than the lenses within the same galaxy.  This trend is likely to be the result of the identification of the bulge components  having the steeper light-profile, characteristic to higher values for the S\'ersic index (see Section~\ref{sec:no_components}). In the case of CCC~158, both the bulge and the lens have similar S\'ersic indices, whereas in CCC~43 the bulge has a significantly lower S\'ersic index. In both cases however, the bulge was selected to be the more extended and less elliptical component.

\begin{figure}
 \includegraphics[width=\columnwidth]{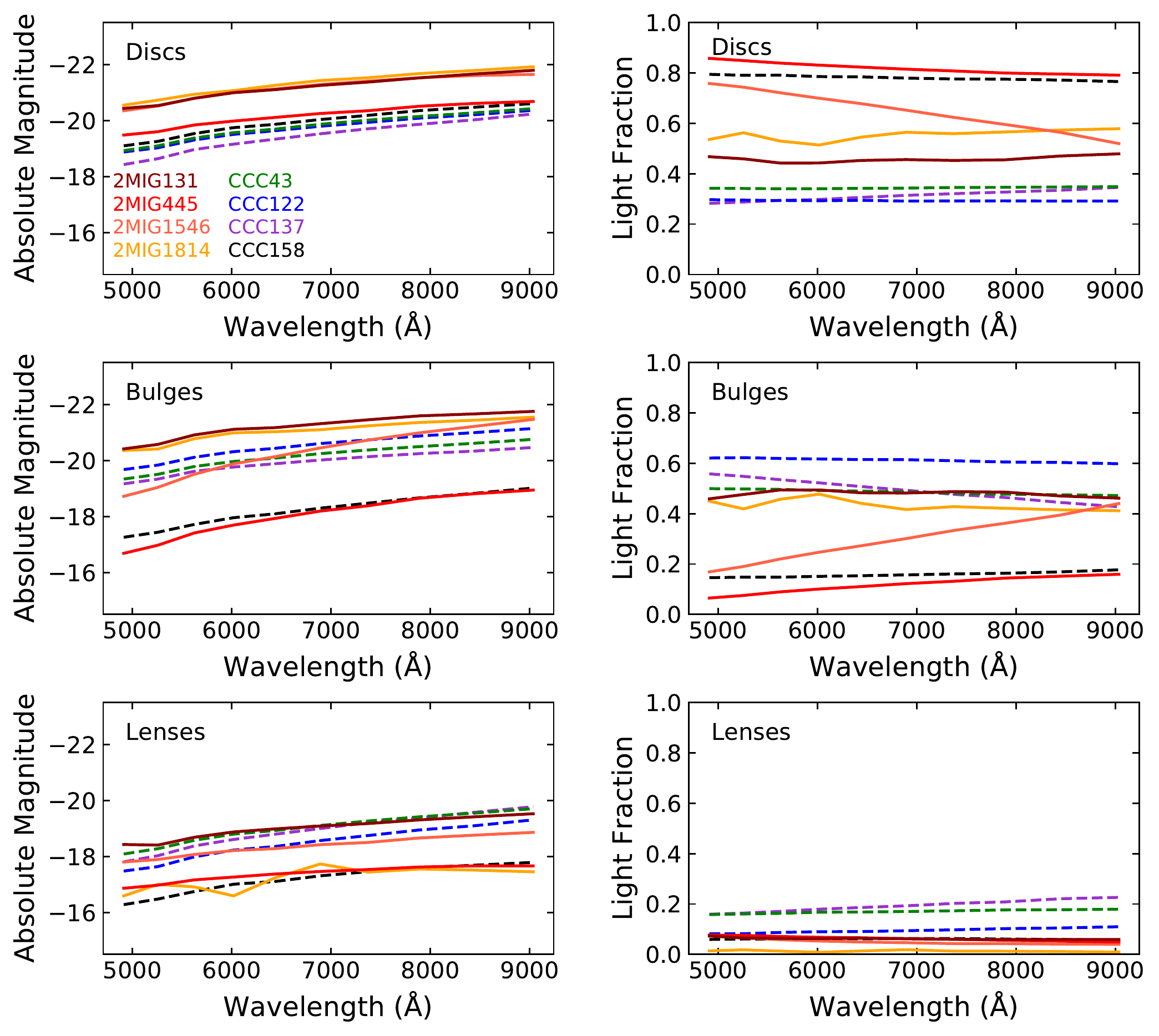}
 \caption{\textit{Left}: The integrated absolute AB magnitudes from the narrow-band images as a function of wavelength for the discs (top), bulges (middle) and lenses (bottom). The cluster galaxies are marked with dashed lines while the isolated galaxies have solid lines. \textit{Right}: The light fraction of each component relative to the total light from the galaxy as a function of wavelength. 
}
 \label{fig:params}
\end{figure}

The discs all have S\'ersic indices $n\lesssim2.5$, which is consistent with the fits to disc galaxies in the Coma Cluster  by  \citet{Weinzirl_2014} and \citet{Head_2015}, and with the models of \citet{Dutton_2009}, in which they found that galaxy discs are not universally exponential. Another feature that is worth mentioning in Fig.~\ref{fig:Re_n} is that the discs of the isolated galaxies have higher S\'ersic indices than those of the cluster galaxies. A grey dashed line has been added to this plot at $n=1.35$ to highlight this result. This trend may indicate that the structure of discs is different in field and cluster S0s, once the contamination from other components has been properly removed. However, with the small sample used in this study, we cannot explore this phenomenon further, nor rule out target selection bias.


\subsection{Colours}\label{sec:colour_grad}
The polynomials used to fit the structural parameters for the fits to the narrow band images, as described in Section~\ref{sec:step_3}, provide information on how those parameters vary as a function of wavelength. From this information, the colours of each component can be derived, which, in turn, can be used to compare their stellar populations. 

The absolute AB magnitudes of each component are plotted against wavelength from the fits to the narrow band images in Fig.~\ref{fig:params}, and clearly show that all the components included in the models to these galaxies are brighter at redder wavelengths. It can be seen that the absolute magnitudes of the discs (top panel) of the cluster galaxies are generally fainter than those of the isolated galaxies. This effect may simply be a selection bias due to the small sample of galaxies and the limited range in their properties. \citetalias{Coccato_2020} showed that the isolated galaxies in this sample have slightly higher masses in general than the cluster galaxies, with mean masses of $\text{log}_{10}(M/M_\odot)\sim11.4$ as opposed to $\sim10.9$, which may explain their brighter discs. The magnitudes of the bulges (middle) and lenses (bottom), on the other hand, show no real correlation with environment. 

The luminosity fractions of each component relative to the total integrated luminosity of the galaxies are also plotted in the middle panel of Fig.~\ref{fig:params}, and show that in all galaxies, the lenses are fainter than their corresponding bulges and discs, with luminosity fractions of $\lesssim0.2$. The light fractions of the lenses are generally consistent with the light fractions of bars in S0s in the Coma Cluster, where \citet{Lansbury_2014} found light fractions of  $\lesssim30\%$. Interestingly, the lenses within the cluster galaxies appear to show higher luminosity fractions than those in the isolated S0s, particularly at redder wavelengths, though again, this trend may be a selection effect.  Additionally, the bulges and discs in many galaxies have relatively flat luminosity fractions with wavelength. This trend may indicate that the bulges and discs have similar SEDs, which could result from poor fits where the bulge and disc have been improperly fitted or where a single component has been modelled with two S\'ersic profiles. In such cases, one would expect to see similar structural parameters for both components. However, upon checking the parameters and the images of each component included in the model (see Appendix~\ref{sec:Appendix_A}), this scenario could be ruled out, and it is therefore more likely that both the bulge and disc in these galaxies have similar colours.

In two of the the isolated galaxies, the bulges are more luminous in the red while the discs are correspondingly fainter in the red and brighter in the blue. These trends may indicate more recent star formation in the discs or higher metallicities in the bulges of those galaxies. Consequently, they may hint at a dependence on the environment or the total mass of the galaxy, but with the small number of galaxies within this sample we cannot rule out that this effect is simply a coincidence.

\subsection{Residual features}\label{sec:residuals}
\label{sec:residuals}

Another benefit of using \textsc{buddi} to model these galaxies is that a residuals datacube is created by subtracting the best fit model created by \textsc{GalfitM} from each image slice. While in  each individual image slice the residuals are almost negligible,  the white-light image  created from this datacube reveals faint structures that would otherwise be hidden in the data. These white-light images created from the residual datacubes are shown for each galaxy in Fig.~\ref{fig:resid_image}, having been scaled to highlight these features, and as part of Appendix~\ref{sec:Appendix_A}, using the same flux scaling as the white-light image for the same galaxy to allow a better comparison of the relative luminosity of these structures.  These images reveal faint structures along the line of sight of each galaxy, including brighter globular clusters within the discs and background galaxies, and thus show the potential to use \textsc{buddi} to cleanly model the light from foreground objects and cleanly extract the spectra of more distant objects within the field.

\begin{figure}
 \includegraphics[width=\linewidth]{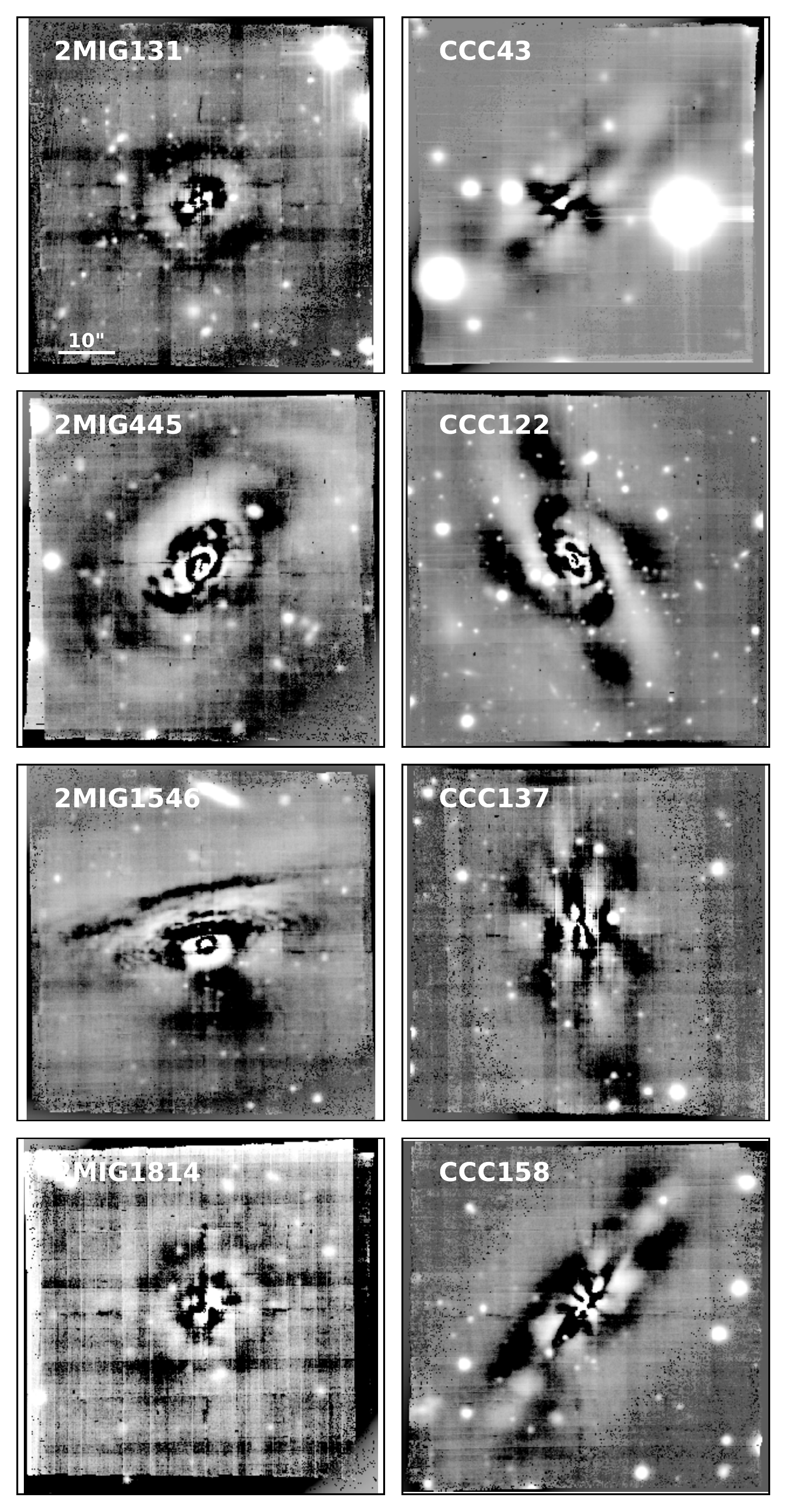}
 \caption{The residual white-light images for the isolated (left) and cluster (right) galaxies, showing the remnants of tidal tails, dusty discs, and spiral arms. The images have been scaled to best display these structures. }
 \label{fig:resid_image}
\end{figure}

One can also see that all galaxies within this sample show evidence of faint features within the galaxies themselves, such as dust lanes or remnants of spiral arms, the latter of which are not visible in the unsharp-mask images in Fig.~\ref{fig:unsharp_mask} nor in the datacubes and white-light images. These spiral arms are particularly prominent in both the inner and outer regions of CCC~122. A comparison with the components used to model this galaxy in Appendix~\ref{sec:Appendix_A} shows that the inner spiral arms lie within the component identified as the lens. While no clear spiral arms were detected in the unsharp-mask image of this galaxy in Fig.~\ref{fig:unsharp_mask}, the inner region of this galaxy does appear to have distinct kinematics in Fig.~\ref{fig:v_sig}. Consequently, it is still possible that the lens component in this galaxy is actually an inner disc, based on the definition used by \citet{Laurikainen_2005}. CCC~158 on the other hand also shows evidence of distinct kinematics in the lens component and  a possible inner edge-on disc in the unsharp-mask image, but no clear spiral structure is detected in the residual image for this galaxy. However, the residual image for this galaxy does contain a lot of structure, which may be created by dust extinction or spiral arms in the outer disc seen edge-on, which may act to hide spiral arms within the lens if present.

The faint spiral arms that can be seen in some of the residual images, particularly for the cluster galaxies, could be evidence of their transformation from spiral galaxies and provide a link between passive spirals and S0s \citep[e.g.][]{Fraser_2018a,Pak_2019}. The residual features in the isolated galaxies, on the other hand, appear more asymmetric, which may reflect different formation processes in each environment. These findings are consistent with those of Paper 1, where the cluster galaxies are more rotationally supported and are thought to have formed through gas removal, while the isolated galaxies are more dispersion supported, which is thought to reflect a transformation more dominated by minor mergers which would disrupt the spiral structure. Attempts were made to try to extract the spectra from the brighter features in the residual datacubes in order to study the different stellar populations across the arms and inter-arm regions, but the faintness of these structures and the consequent low S/N in the extracted spectra made the subsequent analysis unreliable.  

Of particular interest is the residual white-light image for 2MIG~445, where the faint features appear to be remnants of shells or tidal tails following a merger. No clear remnant of the infalling galaxy can be seen, suggesting that if these features were created during a merger, it occurred long enough ago that the merging galaxy has been disrupted completely, while sufficiently recently that the tidal tails have not yet dissipated entirely \citep[$\sim1$--$2\,$Gyr ago, cf.][]{Eliche_2018}. Alternatively, the merger could have been minor, thus preventing significant disruption to the structure of the galaxy. While the stellar population maps for this galaxy show no clear trends that could be interpreted as the differences between those of the progenitor and infalling galaxies, the kinematics maps in Fig.~A1 of \citetalias{Coccato_2020} show a clear distortion to the North-West of the centre of the galaxy, particularly in the velocity dispersion and $h_4$ parameters. No clear optical counterpart can be seen in Figures~\ref{fig:unsharp_mask} and \ref{fig:resid_image}, thus indicating that this disturbance in the kinematics may be a remnant of a merger where the infalling galaxy has been destroyed and the kinematics progenitor galaxy are still somewhat disrupted.


\section{Stellar populations in the bulges, discs and lenses}\label{sec:stellar_pops}

\begin{figure*}
 \includegraphics[width=0.33\linewidth]{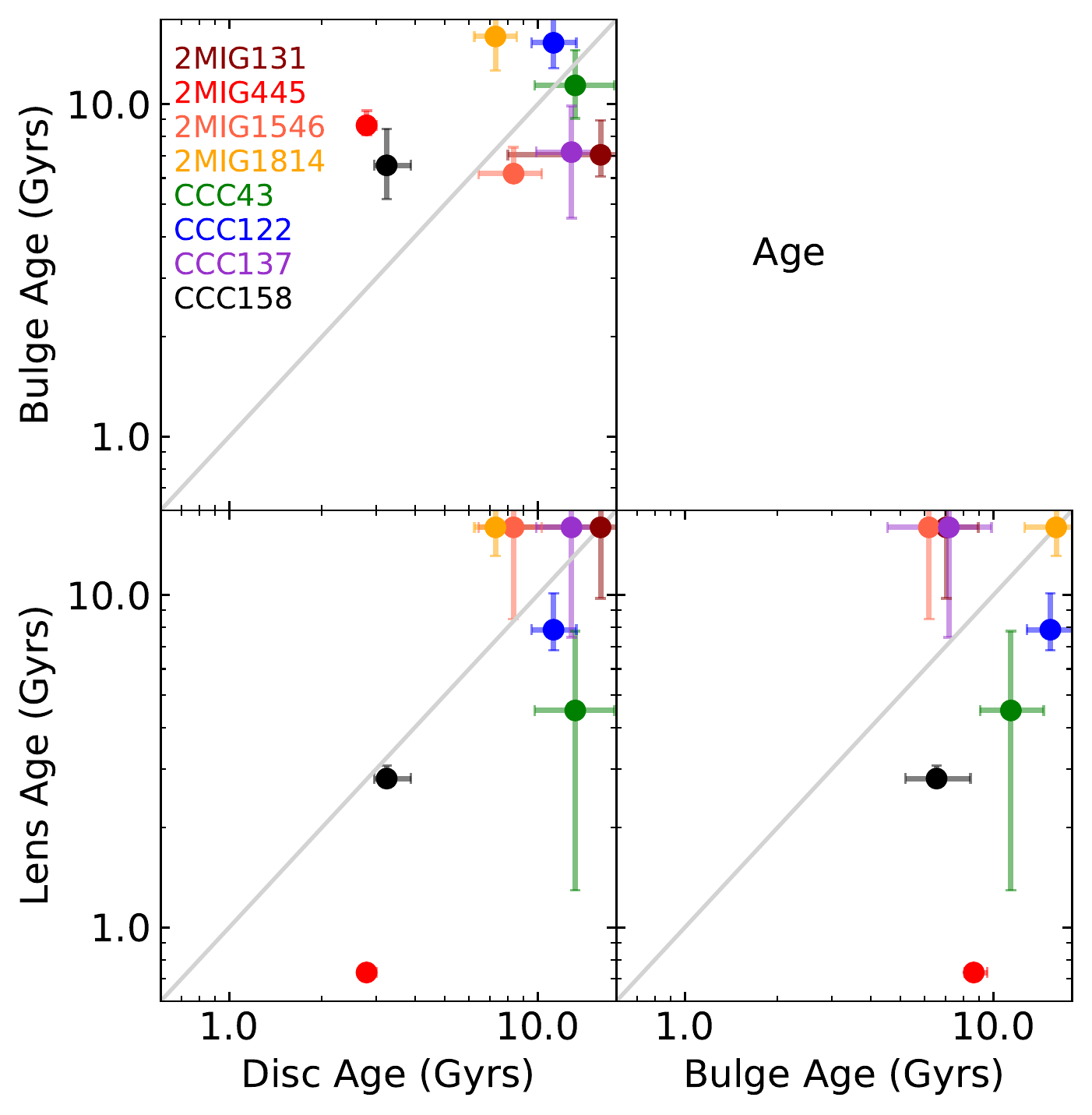}
 \includegraphics[width=0.33\linewidth]{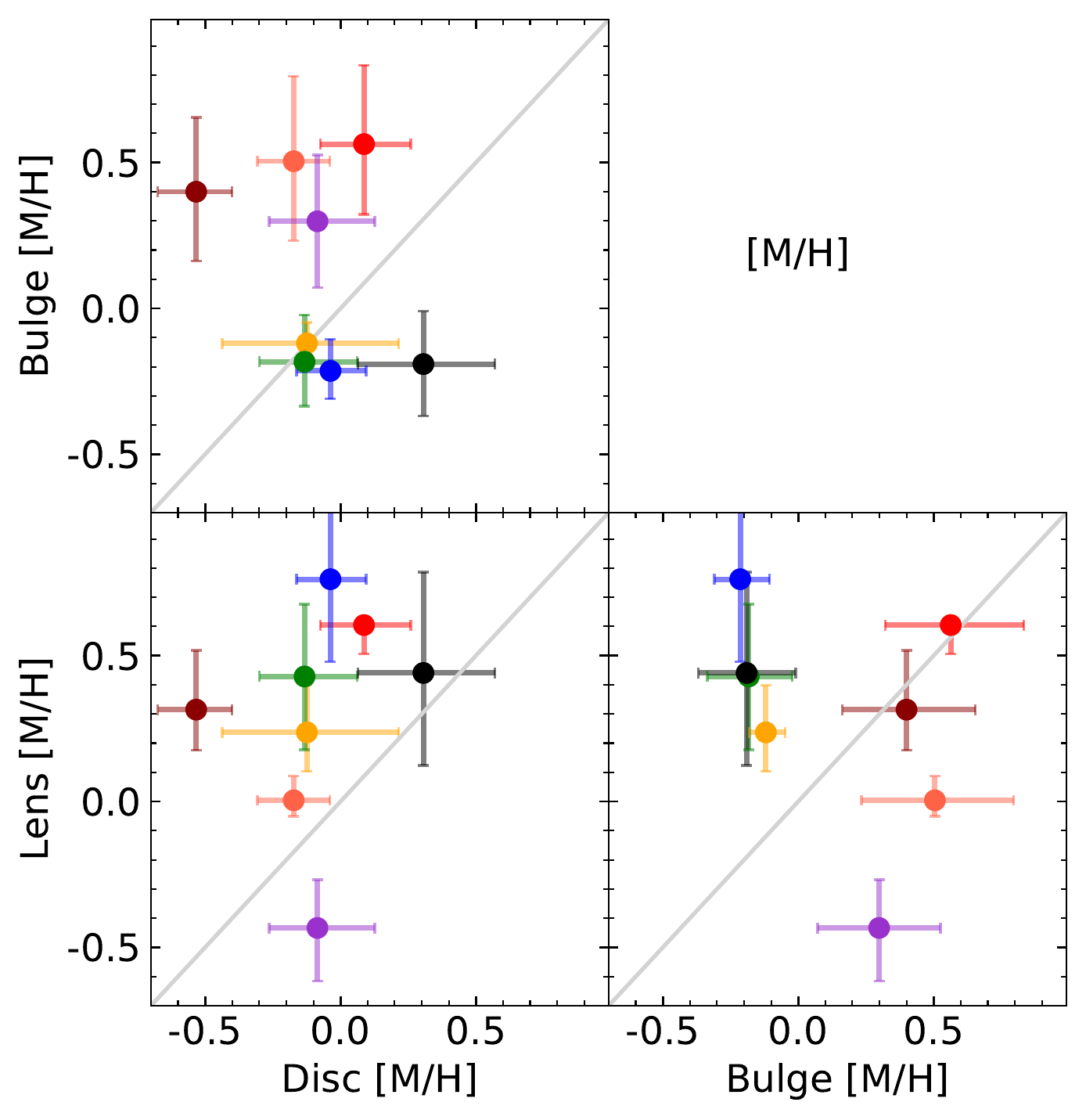}
 \includegraphics[width=0.33\linewidth]{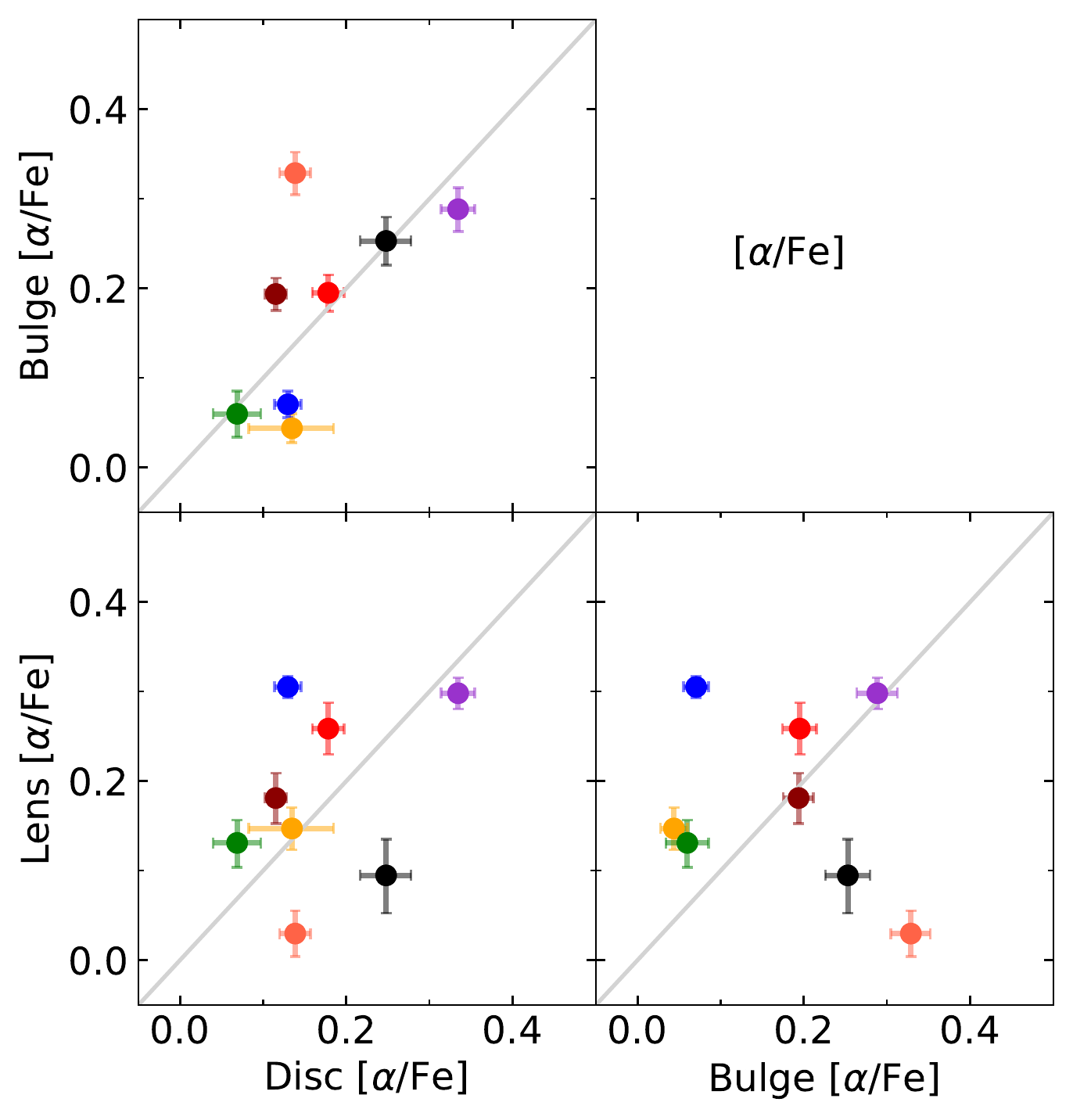}
 \caption{A comparison of the luminosity-weighted ages (left), metallicities (middle) and $\alpha$-enhancement (right) between the bulges, discs and lenses of each galaxy }
 \label{fig:LW_stellar_pops}
\end{figure*}

\begin{figure*}
 \includegraphics[width=0.33\linewidth]{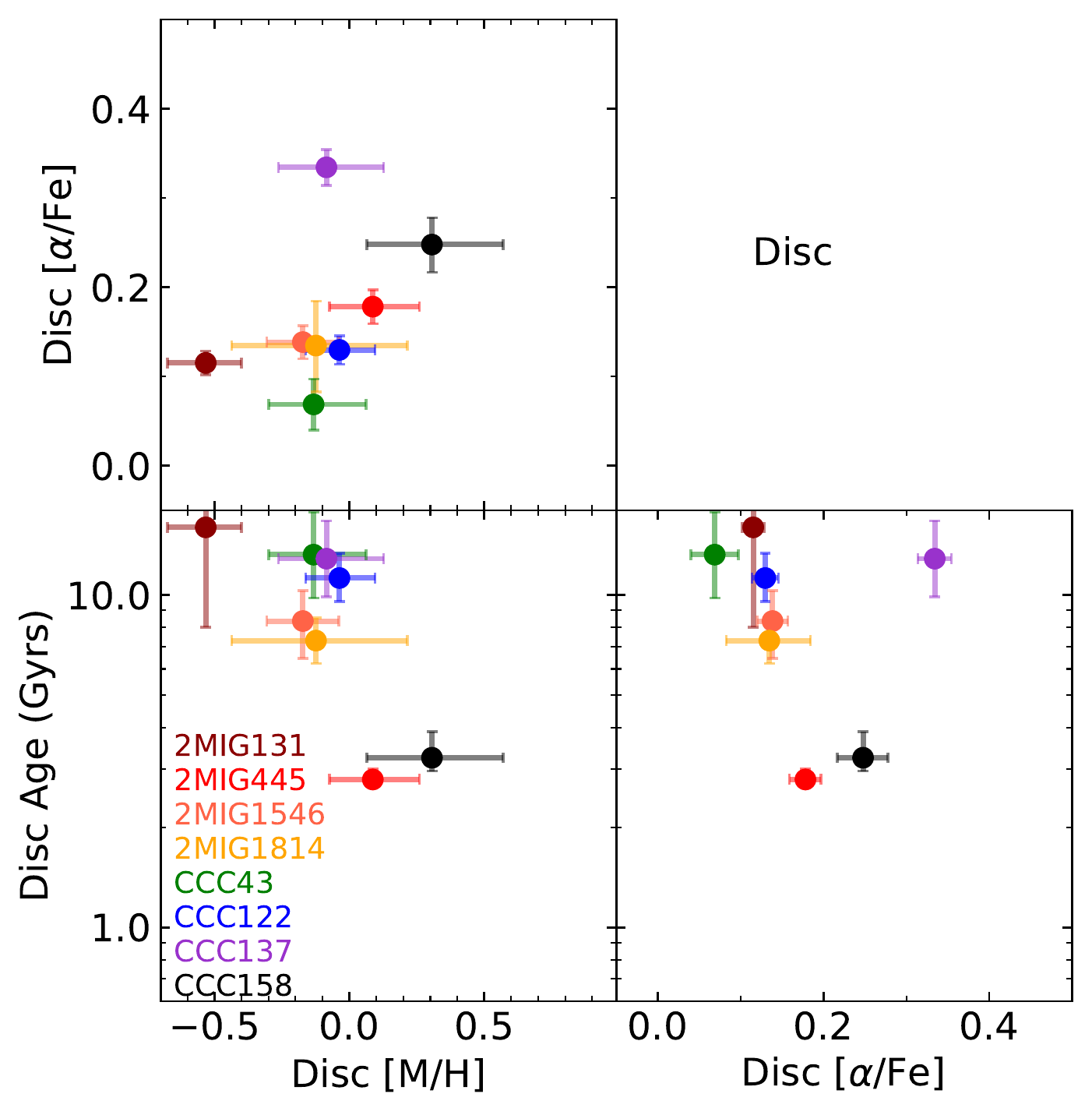}
 \includegraphics[width=0.33\linewidth]{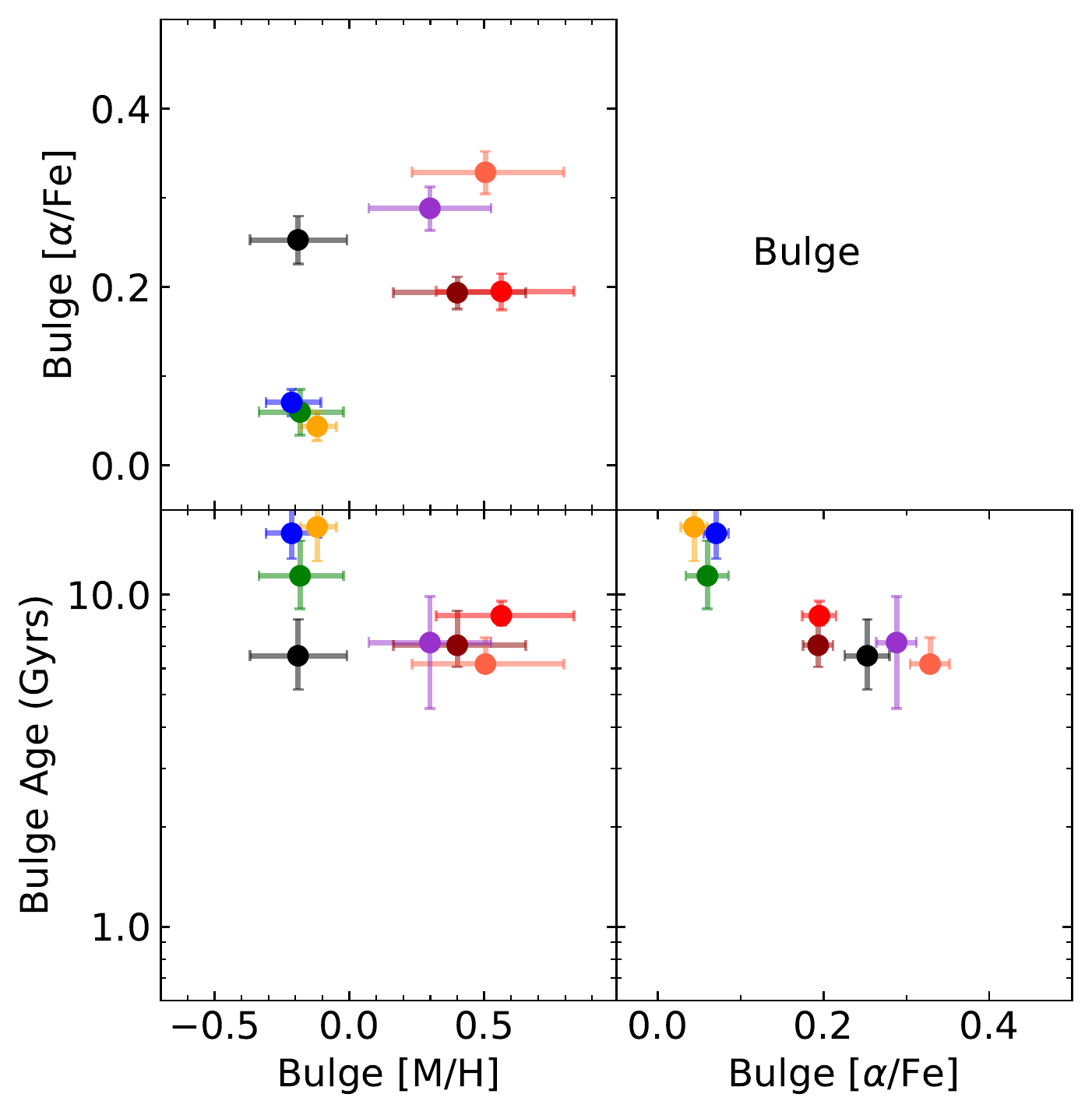}
 \includegraphics[width=0.33\linewidth]{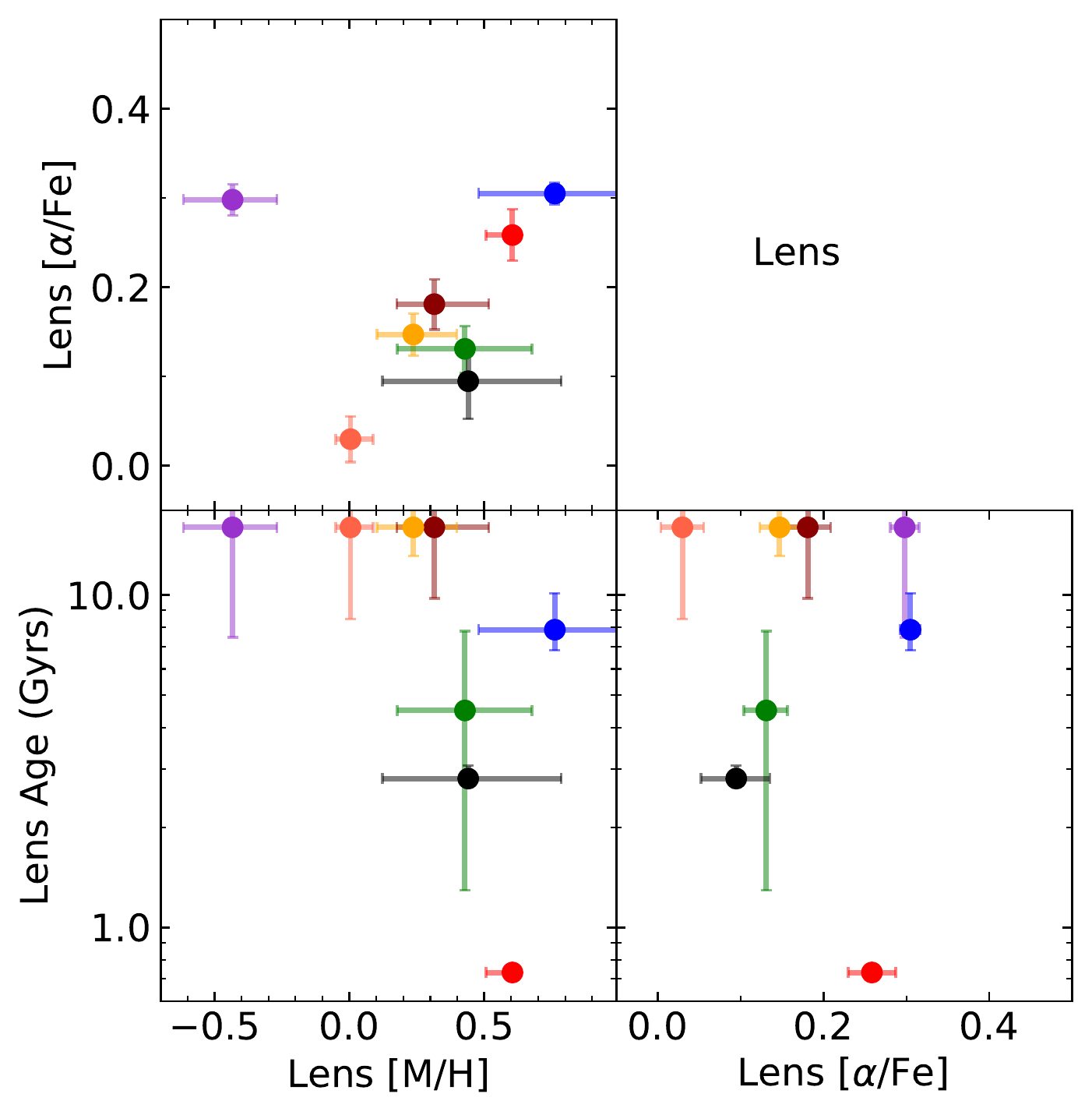}
 \caption{A comparison of the luminosity-weighted ages, metallicities and $\alpha$-enhancement between the discs (left), bulges (middle) and lenses (right) of each galaxy }
 \label{fig:LW_stellar_pops2}
\end{figure*}

\subsection{Luminosity-Weighted Stellar Populations}\label{sec:LW_pops}
Having extracted the spectra for the bulges, discs and lenses within each galaxy, estimates for their luminosity-weighted stellar populations were derived using the same techniques outlined in Sections~\ref{sec:LW age met} and \ref{sec:LW alpha}. The results are given in Fig.~\ref{fig:LW_stellar_pops}, showing the correlations between the ages, metallicities and $\alpha$-enhancements between the discs, bulges and lenses, and  Fig.~\ref{fig:LW_stellar_pops2}, which presents the trends between these properties for each component separately. The spectra for the PSF components were found to have very low S/N and their line strength measurements fell significantly off the grids in the direction of old, metal-rich stellar populations. As a result, they have been omitted from the stellar populations analysis presented in this work.

It can be seen that with the exception of 2MIG~445 and CCC~158 all of the discs are generally old, with ages~$>8\,$Gyr, while the bulges are all older than $\sim6\,$Gyr and the lenses show a wide range of ages. Due to the old ages of the components, only large differences in the ages (i.e. several Gyr) between each component would be detected in these plots. Consequently with the large uncertainties in these age measurements, no clear trend can be seen between the ages of the discs with either those of the bulges or lenses or as a function of environment. This result is consistent with the relatively flat age gradients and generally old ages for all the galaxies in Figures~\ref{fig:radial_stellar_pops_maps} and \ref{fig:radial_stellar_pops_plots}. However, the lenses of 2MIG~445, CCC~43 and CCC~158 do show evidence of significantly younger stellar populations, indicating that these components hosted the most recent star formation within these galaxies. 

The results for the metallicities  are more interesting, showing that the metallicities of the lenses are generally consistent with or higher than the discs, with the exception of CCC~137, which may reflect that the lenses have formed in situ from the disc material. Furthermore, with the exception of CCC~158, the bulges generally display consistent or higher metallicities than the discs as well, though no clear trend is seen between the metallicities of the lenses and bulges. 
These higher metallicities in the more compact components compared to the discs is similar to the findings of  \citet{Johnston_2012,Johnston_2014} that the bulges of S0s in the Virgo and Fornax Clusters generally contain more metal-rich stellar populations, though the younger stellar populations in the same regions is not replicated  in this work. This difference in results may be due to the increased information available in IFU spectroscopy compared to long-slit spectroscopy, or due to the additional components that can be more reliably modelled with this data set. Figures~\ref{fig:radial_stellar_pops_maps} and \ref{fig:radial_stellar_pops_plots} also showed higher metallicities in the inner regions of all the galaxies, and so it now appears likely that these gradients are a result of the superposition of varying fractions of light from the bulges, discs and lenses at each point in the galaxy. For example, the younger, more metal-rich stellar populations detected at small radii CCC~43, CCC~158 and 2MIG~445 in Figures~\ref{fig:radial_stellar_pops_maps}  and \ref{fig:radial_stellar_pops_plots} can be attributed to the lens component in Fig.~\ref{fig:LW_stellar_pops}. Similarly, the uniformly old stellar populations measured at all radii in 2MIG~1814 and  CCC~122 in Figures~\ref{fig:radial_stellar_pops_maps}  and \ref{fig:radial_stellar_pops_plots} are also reflected in the ages of the individual components. Together, these plots show the strength of \textsc{buddi} at extracting stellar populations parameters for individual structures within these galaxies despite using no information on the stellar populations in the fits.

In the plots for the $\alpha$-enhancement, it can be seen that the [$\alpha$/Fe] ratio for the bulges are correlated with those for the discs, indicating that the most recent star formation activity in both components were connected. Only 2MIG~1546 differs, with a higher $\alpha$-enhancement in the bulge than the disc. A similar trend between the bulge and disc $\alpha$-enhancement was seen in \citet{Johnston_2014} for S0s in Virgo, in which it was attributed to a final episode of star formation within the bulge, fuelled by gas that originated in the disc, which took place after the star formation in the disc had ceased. However, in that study the bulges were found to contain younger, more metal-rich stars than their discs, which is not the case in this sample. Instead, almost all the bulges and discs show evidence of old stellar populations, with the differences in their ages being smaller or similar to the uncertainties in the estimates. Consequently, the consistent [$\alpha$/Fe] ratios between these two components suggests that they also formed over similar timescales. 

The [$\alpha$/Fe] ratios of the lens components on the other hand show no clear correlation with either the bulges or discs, indicating that they were formed through a distinct and more random star-formation episode. The metallicities of the lenses however are generally consistent with or higher than the discs, with the exception of CCC~137, which may reflect that the lenses have formed from the disc material. This scenario therefore rules out the theory that the lens components are simply artefacts resulting from a poor bulge+disc fit to S0 galaxies.


\subsection{Mass-Weighted Stellar Populations}\label{sec:MW_stellar_pops}
The stellar populations derived through analysis of line strengths only provides the luminosity-weighted properties, which can be dominated by stars produced in a recent star formation event and therefore may not reflect the true age of an old galaxy. Consequently, they are best used to determine how long ago the most recent star formation activity occurred. However, through full spectral fitting using template spectra for a wide range of ages and metallicities, it is  possible to obtain estimates of the mass-weighted stellar populations and star-formation histories for each component within these galaxies to better understand their relationship.

\begin{figure*}
 \includegraphics[width=0.99\linewidth]{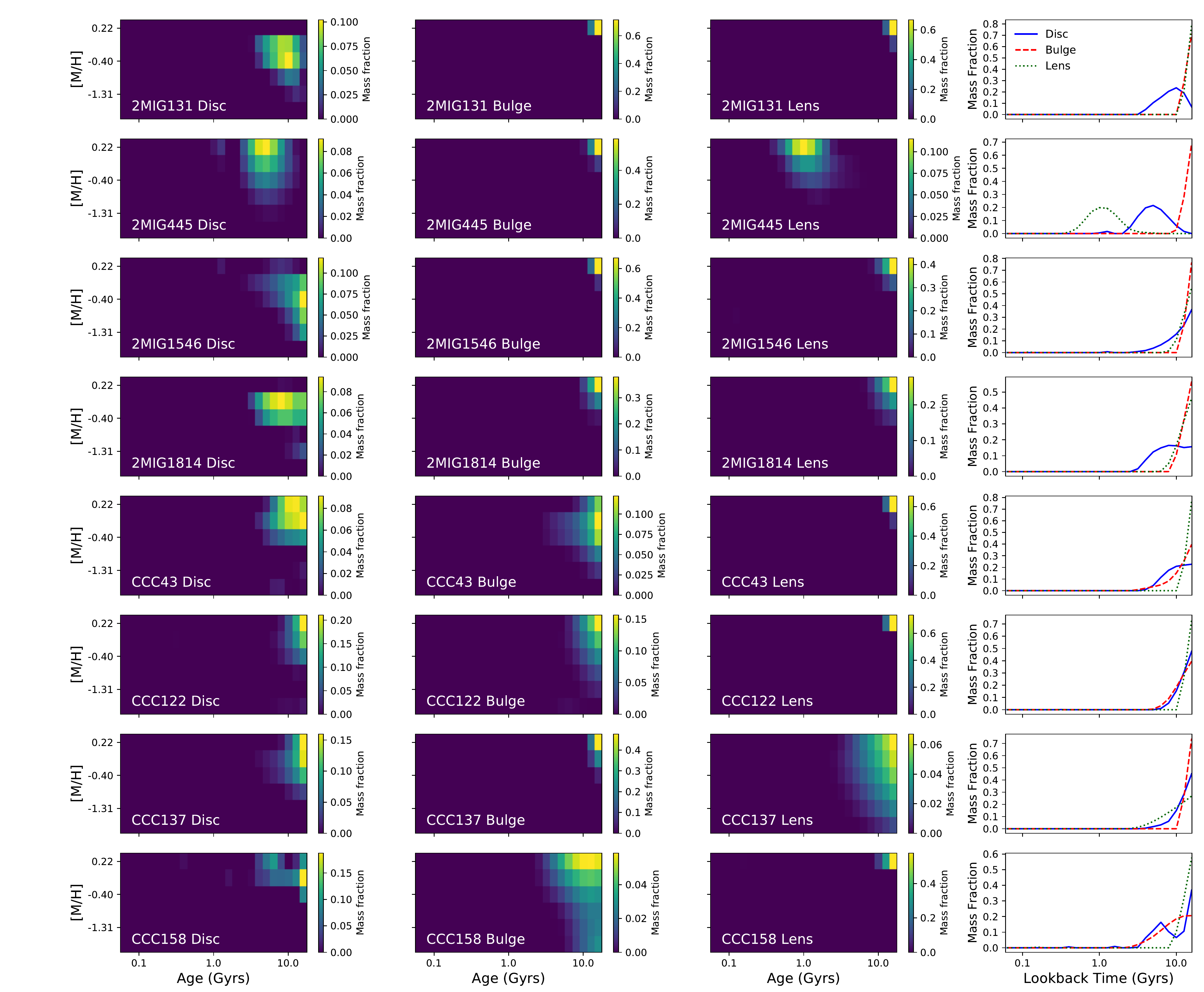}
 \caption{The star formation histories of each component, displayed as the mass fraction created with each look-back time and metallicity included within the model grids for the discs, bulges and lenses (first to third columns respectively), and as the mass fraction created as a function of lookback time (right).}
 \label{fig:weights}
\end{figure*}

The mass-weighted ages and metallicities were derived from the spectrum of each component using \textsc{ppxf} to apply a regularized fit to the spectrum. A multiplicative Legendre polynomial of order 10 was also included in the fit to correct the shape of the continuum, reducing the sensitivity to dust reddening and omitting the requirement of a reddening curve \citep{Cappellari_2017}. The fit used a linear combination of template spectra of known relative ages and metallicities from the MILES library spanning an age and metallicity range of $0.06-18\,$Gyr and  [M/H]=$-2.32$ to +0.22 respectively.  The regularization process within \textsc{ppxf} works to create a smoothed variation in the weights of the templates with similar ages and metallicities, where the degree of smoothing is controlled by the regularization parameter determined by the user. The process started with an unregularized fit to each spectrum to measure the $\chi^2$ value for the fit, and then the noise spectrum was scaled appropriately until $\chi^2/N_{DOF} = 1$, where $N_{DOF}$ is the number of degrees of freedom in the fit, which corresponds to the number of unmasked pixels in the input spectrum. The fit to the spectrum was then repeated using this scaled noise spectrum and increasing values for the regularization parameter until the $\chi^2$ of the fit increased by $\Delta\chi^2 = \sqrt{2\times N_{DOF}}$. This value represents the limit between a smooth fit that still reflects the star-formation history of the galaxy and one that has been smoothed excessively. It should be noted at this point that this smoothed fit may not  reflect the true star-formation history of the galaxy, which is likely to vary over shorter timescales than the models allow, but instead acts to reduce the age-metallicity degeneracy between spectra, thus allowing a more consistent comparison of systematic trends in their star-formation histories.

Since the template spectra are modelled for an initial birth cloud mass of 1~Solar mass, the final weight of each template reflects the `zero-age' mass-to-light ratio of that stellar population. Consequently, the smoothed variation in the weights of neighbouring template spectra in the final regularised fit represents a simplified star-formation history of that part of the galaxy by defining the relative mass contribution of each stellar population \citep[see e.g.][]{McDermid_2015}. These star-formation histories for each component are given in Fig.~\ref{fig:weights}, showing both the mass fraction associated to each step in age and metallicity within the models, and the mass fraction created as a function of lookback time. It is immediately clear in these plots that all the components are old and metal-rich, with only the disc and lens of 2MIG~445 and the bulge and disc of CCC~158 showing evidence of significant recent star formation. These results are in agreement with the luminosity-weighted stellar populations, which show evidence of younger stellar populations within these same components. However, these plots for the star-formation histories go beyond to show that all of the galaxies in this sample have underlying old, metal-rich stellar populations, indicating that the majority of their mass formed long ago and that any younger stellar populations detected represent a later star-formation episode that contributed very little towards the mass of that component.

\begin{figure}
 \includegraphics[width=0.8\columnwidth]{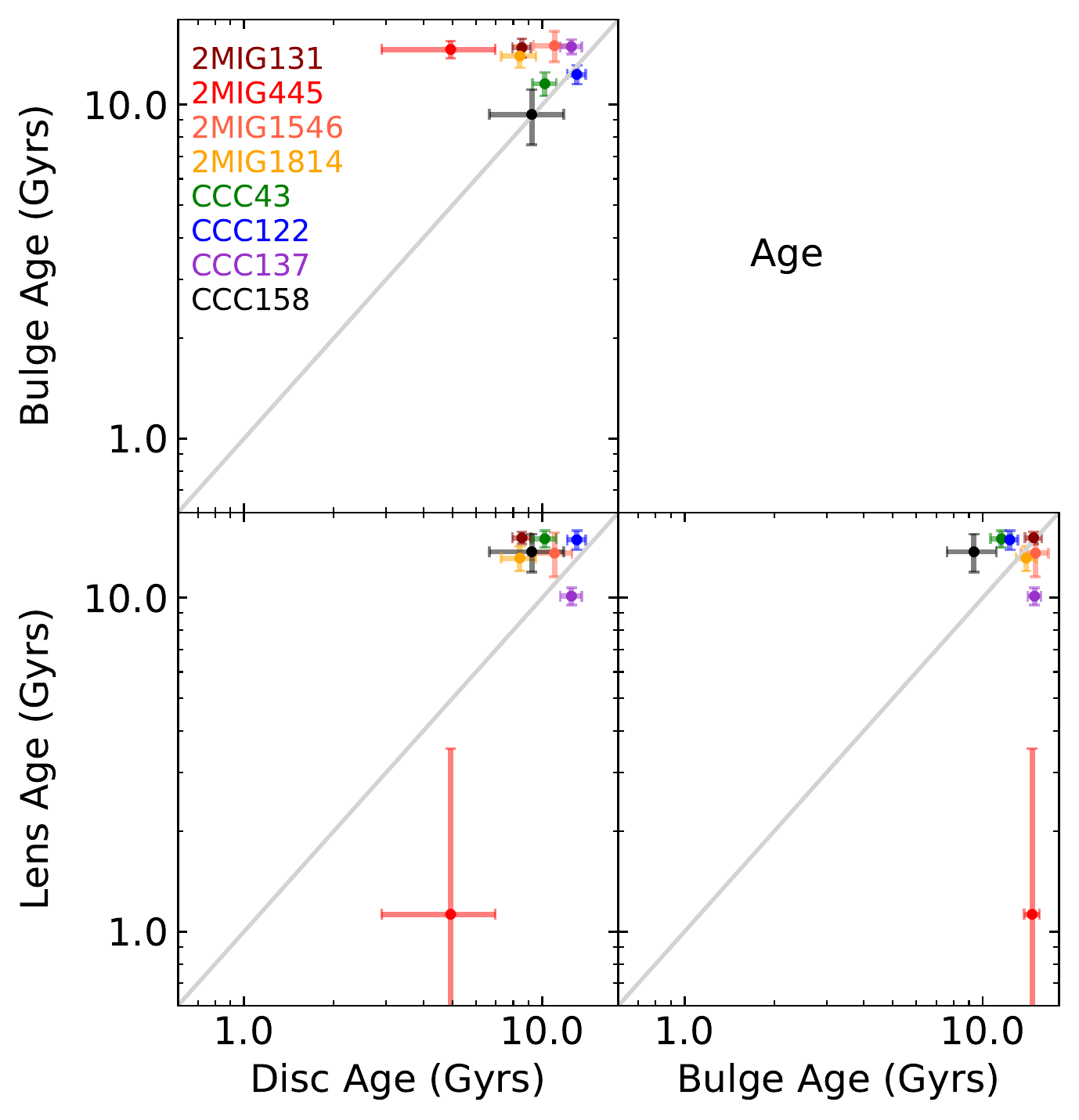}
 \includegraphics[width=0.8\columnwidth]{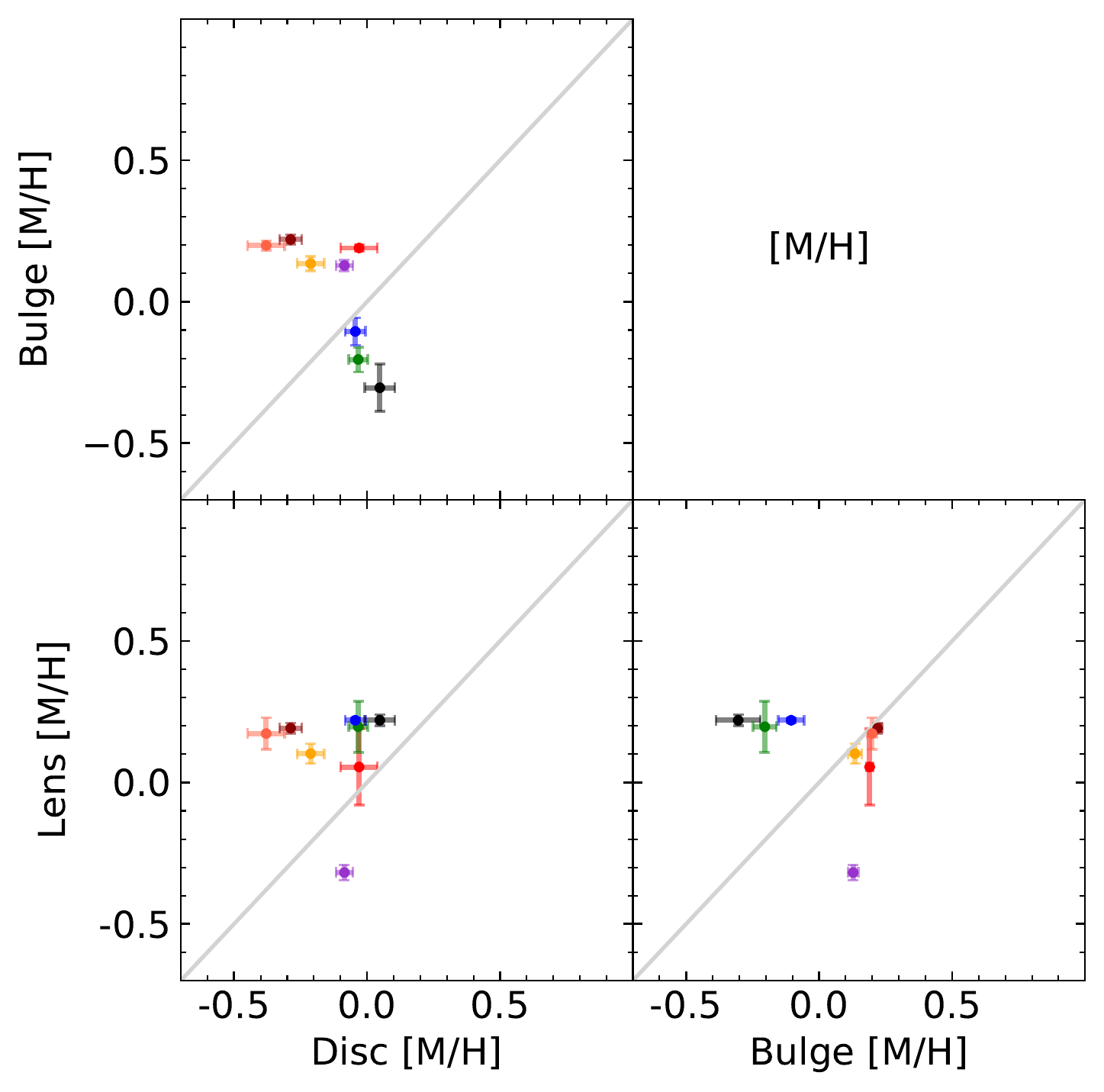}
 \caption{Mass-weighted stellar populations, showing the relative mass-weighted ages (top) and metallicities (bottom) of the discs, bulges and lenses.}
 \label{fig:MW_stellar_pops}
\end{figure}

The mean values for the mass-weighted ages and metallicities of each component can be calculated from the weights of the template spectra using 
\begin{equation} 
	\text{log(Age$_{\text{M-W}}$)}=\frac{\sum \omega_{i} \text{log(Age$_{\text{template},i}$)}}{\sum \omega_{i}}
	\label{eq:age}
\end{equation}
and 
\begin{equation} 
	\text{[M/H]$_{\text{M-W}}$}=\frac{\sum \omega_{i} \text{[M/H]}_{\text{template},i}}{\sum \omega_{i}}
	\label{eq:met}
\end{equation}
respectively, where $\omega_{i}$ represents the weight of the $i^{th}$ template (i.e. the value by which the $i^{th}$ template stellar template is multiplied to best fit the galaxy spectrum), and [M/H]$_{\text{template},i}$ and Age$_{\text{template},i}$ are the metallicity and age of the $i^{th}$ template respectively. These results are plotted in Fig.~\ref{fig:MW_stellar_pops}, and the associated uncertainties were calculated through a series of simulations in which random noise added to the best fit model to each spectrum to replicate the S/N of the original spectrum. One hundred simulated spectra were created for each component in this way, and the mass-weighted ages and metallicities were measured from these spectra using the same values for the regularization. The final uncertainties were calculated as the standard deviation in the measurements from these simulations.

In general, these results also reflect that the majority of the mass within these structures is made up of of old, metal-rich stellar populations, with only the lens and disc of 2MIG~445  showing evidence of a significant fraction of young stars. 

It is possible to see in Fig.~\ref{fig:MW_stellar_pops} that the mass-weighted metallicities of the bulges and lenses of the isolated galaxies show much smaller distribution than the same components in the cluster galaxies, whereas the opposite is true for the disc components. Additionally, the bulges of all the isolated galaxies and CCC137 again show higher metallicities than the discs, while the bulge of 2MIG~445 is significantly older than the disc. These stellar populations could also explain the redder bulges and bluer discs in isolated galaxies compared to the cluster galaxies in Fig.~\ref{fig:params}. However, the low numbers of galaxies from each environment in this study makes it difficult to determine if this trend is true or coincidence, and it will be followed up in a future work with a larger sample of galaxies.


\section{Discussion and Conclusions}
\label{sec:Discussion}

We have carried out a detailed analysis of the stellar populations and star-formation histories of the independent components within a sample of unbarred S0s galaxies in isolated and cluster environments.  Not only were the models of these galaxies expanded beyond the typical assumptions of an exponential disc surrounding a S\'ersic or de~Vaucouleurs bulge, but we also incorporated the first spectroscopic analysis of lenses within S0s, based on our search of the literature.  Through analysis of the spectra for each component, we derived estimates of their stellar populations in order to determine their contributions towards the evolution of the galaxies as a whole. Using \textsc{buddi}, the light profiles of each galaxy were modelled in each image slice of the MUSE observations in order to cleanly model the luminosity and extract the spectrum of each component, allowing measurements of their stellar populations with minimal contamination.

The light profile fits revealed that all eight galaxies required more complex models than a simple bulge+disc fit, consisting of three extended components in all cases with a further three galaxies requiring a PSF profile at their cores as well. In each case, the disc was identified by eye as the most extended component, while the bulge was distinguished as the inner spheroidal component with the steeper surface brightness profile and softer edge, which in the majority of the sample also corresponded to the brightest component at the core of the galaxy. In each galaxy, the third extended component satisfied the criteria for being a lens, having a flat surface brightness profile with a sharp outer edge \citep{Kormendy_1979}, but they did display a range of elongations, with axes ratios ranging from 0.41 to 0.96. The kinematics, unsharp mask images and stellar populations maps of each galaxy indicated that this component may instead be an inner edge-on disc in CCC~158, but without conclusive evidence it has been included it under the lens umbrella for the purpose of this work. 

The stellar populations within each component were analysed through both line strength measurements and regularized full spectral fitting to obtain estimates of the luminosity and mass-weighted properties respectively. It was found that in general, the bulges and discs of all galaxies are very old, with only the discs of 2MIG~445 and CCC~158 showing significantly younger stellar populations. This result is inconsistent with the findings of \citet{Johnston_2012,Johnston_2014}, where the bulges of S0s in Virgo and Fornax were found to have younger and more metal-rich stellar populations than their discs. This difference could be a result of the different evolutionary pathways of the samples of galaxies used, but may also be an artefact from the fits to the one-dimensional light profile in the long-slit spectra in that work and the inclusion of the lens component in this work. \citet{Laurikainen_2009} found that modelling the two-dimensional light profile of a galaxy gave superior results than using the one dimensional light profile, particularly when the galaxy contains an additional component beyond the bulge and disc. However, the older bulges detected in this study are consistent with other studies of S0s. For example, \citet{Fraser_2018b} found that the bulges in MaNGA S0s with similar masses to those used in this study are generally older than their discs, and \citet{Katkov_2014,Katkov_2015}  identified that S0s in isolated environments tend to have similar ages in both the bulges and discs. \citet{Mendez_2019b} also found that the bulges in S0s are older than their discs, which they interpreted as evidence that star formation occurs throughout the disc, and only in the disc, not in the bulge.  Another study by \citet{Rodriguez_2014} of cluster `k+A' galaxies, which are thought to be an intermediate phase in the transformation of spirals to S0s, found that the final episode of star formation was located within the extended disc, though more centrally concentrated than the earlier star formation.

The grids showing the mass fractions created as a function of time and metallicity further expand on this result, showing that nearly all the bulges and discs have an underlying high mass of intermediate to old stars, with some more recent star formation occurring that dominates the light in these cases but only contributes towards a small fraction of the mass of that component. In general, it can be seen that the majority of the mass in the discs was built up over a longer timescale than in the bulges, which is consistent with  spiral galaxies, which have star-forming discs and red bulges, being the progenitors of these S0s. However, in CCC~122 and CCC~137 the star-formation timescales in the bulges and discs appear to be similar, possibly indicating that the star formation was truncated soon after the progenitor spiral was formed. With the exception of CCC~158, the luminosity and mass-weighted metallicities of the bulges are all consistent with or higher than those of the discs, indicating that they formed from the same material. Furthermore, the [$\alpha$/Fe] of the bulges and discs show a strong correlation, indicating that their most recent episodes of star formation are somehow connected. \citet{Morelli_2008} and \citet{Morelli_2016} studied the stellar populations within bulges of cluster and isolated S0s respectively, finding that galaxies in both environments display null age and negative metallicity gradients with radius, and that the isolated galaxies show negative [$\alpha$/Fe] gradients while the cluster galaxies show no slope in this property. With these results they concluded that the bulges all formed through gas dissipating inwards and fuelling star formation in the inner regions of the galaxy, with successive mergers only playing a significant role in the cluster galaxies, where they have flattened out the gradients in [$\alpha$/Fe]. On the other hand, \citet{Katkov_2015} proposed that the similar [$\alpha$/Fe] ratios in the bulges and discs instead point to a scenario where star formation in both components started simultaneously and ceased at around the same time. Furthermore, \citet{Fraser_2018b} identified that the process that truncated the star formation is more dependent on the galaxy mass than environment, though their sample of galaxies didn't include S0s from the densest environments, and are best used to compare to the results for the isolated galaxies in this work. Our results for the bulges and discs are consistent with these studies. The  similarly old mass-weighted ages and similar or higher metallicites in the bulges relative to the discs  present a scenario where the majority of the mass in the bulges and discs was created long ago, and that these components have survived without experiencing any major mergers that sufficiently disrupted their discy morphology since then \citep{Prochaska_2011}. The wider range in luminosity-weighted ages however indicate that most galaxies have experience a small amount of more recent star formation after the majority of their mass was built up, with the consistent $\alpha$-enhancements between these components hinting that this most recent episode of star formation occurred over similar timescales in the bulge and disc of each galaxy. However, the youngest components and more extended star-formation histories are observed mainly in the field S0 of our sample (see Fig.~\ref{fig:weights}), suggesting that minor events (e.g. minor mergers) might have occurred after the formation of the main bulk of stars in less dense environments; this scenario is consistent with the conclusions of \citetalias{Coccato_2020}, in which the kinematic properties of field S0s suggest that they have experienced minor mergers. Furthermore, the residual images of 2MIG~445 revealed shells and tidal tails, which would further support this theory.


The lenses were found to have  generally higher metallicities than the discs, indicating that they may have formed in situ from the same material.  However, where the bulges and discs show a strong correlation in the [$\alpha$/Fe] ratios, the lenses show no correlation with either of these components.  Furthermore, the lenses in CCC~43 and CCC~158 showed younger luminosity-weighted stellar populations, while the mass-weighted ages of all the lenses were found to be the same as or older than the discs. The lens of 2MIG~445, however, shows a mass-weighted age of $\sim2\,$Gyr and thus has been formed more recently. Simulations by \citet{Eliche_2012,Eliche_2018} found that the typical timescale for a galaxy to become a relaxed elliptical or S0 following both minor and major mergers is $1-2\,$Gyr, though  \citet{Borlaff_2014} found a timescale of 3.5~Gyr to create S0s with truncated discs. These timescales are comparable to the age measurement for the lens in 2MIG~445, and thus are consistent with the scenario where this galaxy has undergone a recent merger that may have contributed towards the formation of the lens. \citet{Morelli_2008,Morelli_2016} proposed that the flat radial gradients in [$\alpha$/Fe] ratios are produced by successive mergers mixing the stellar populations. Therefore, if true, the lack of correlation in the [$\alpha$/Fe] ratios in the lenses with those of the bulges and discs could mean that the most recent star formation in the lenses occurred after the mergers, while an alternative scenario is that the most recent episode of star formation in the bulges and discs occurred over similar timescales and were fuelled by the same material, while the lenses formed independently over more random timescales within the lifetimes of their host galaxies. 

Another open question about the origin of lenses is their connection to bars. \citet{Kormendy_1979} proposed that lenses in unbarred galaxies are the remnants of dissolved bars that have spilled out into a lens shape. A study by \citet{Laurikainen_2013} found that the  stellar populations derived from photometry of bars and lenses in S0s are similar, which they believe is consistent with this theory that lenses are partially destroyed bars. On the other hand, models by \citet{Athanassoula_1983} showed that  galaxies that are too dynamically hot to form bars may still be cool enough to form lenses through disc instabilities, while \citet{Bosma_1983} found evidence that lenses in unbarred galaxies are redder than their discs, and so likely formed through truncated star formation early in the life of the galaxy. If lenses were connected to bars, one would expect to see similar stellar populations properties in bars and lenses relative to the disc. In the literature, bars have been found to be more metal-rich and less $\alpha$-enhanced than their surrounding discs by \citet{Neumann_2020}, while \citet{Carrillo_2020} found that the bar in NGC~2903 has similar metallicity and a higher fraction of younger stars than the outer disc, which they interpreted as evidence that gas flowing along the bar has triggered more recent star-formation activity. Further evidence for ongoing star-formation along the leading edges of bars in spiral galaxies has also been found by \citet{Peterken_2019} and \citet{Fraser_2020}. In the sample of unbarred S0s used in this  work, we found that the lenses tend to have higher metallicities than discs, while the discs display more recent or longer episodes of star formation than the lenses. While these results are consistent with the theory that lenses are partially dissolved bars, they cannot rule out the scenarios proposed by \citet{Athanassoula_1983} and \citet{Bosma_1983}.

Together, the results presented in this work present a scenario in which the bulges, discs and lenses in these S0 galaxies formed in situ from the same material, but while the most recent star formation activity in the bulges and discs were truncated at around the same time and after similar star formation timescales, the lenses appear to have formed at more random times over the lifetime of the galaxy and over different timescales. One explanation for these results could be that the bulges and discs have formed through dissipational processes, such as major mergers or gas accretion, at high redshift, with later evolution  through accretion of less massive satellites and internal mechanisms induced by a bar. The presence of younger stellar populations, asymmetric features in the residual images and the dispersion supported kinematics reported in \citetalias{Coccato_2020} in the field S0s may indicate that these galaxies have been affected more by minor mergers than the cluster galaxies, but the low numbers used in this study mean that no firm conclusions can be derived as to the factors affecting the formation of the bulges, discs and lenses as a function of environment. However, this scenario is consistent with the theories for the formation of high-mass S0s by \citet{vandenBergh_2009}, \citet{Barway_2013} and \citet{Mendez_2019b}. 
The lack of correlation in the ages and $\alpha$-enhancements of the lenses with the bulges and discs may indicate a later formation scenario for these components from evolved bars, which fits with the scenario that the frequency of bars drops in S0s  while they also appear more evolved in these galaxies \citep{Laurikainen_2009}.

\section*{Acknowledgements}
\addcontentsline{toc}{section}{Acknowledgements}

This study was based on observations collected at the European Organisation for Astronomical Research in the Southern Hemisphere under ESO programme 096.B-0325(A), PI: Jaff\'e. 
E.J.J. acknowledges support from FONDECYT Postdoctoral Fellowship Project No.\,3180557 and the BASAL Center for Astrophysics and Associated Technologies (PFB-06). 
YJ acknowledges financial support from  CONICYT PAI (Concurso Nacional de Inserci\'on en la Academia 2017) No. 79170132, and FONDECYT Iniciaci\'on 2018 No. 11180558. 
BRP acknowledges support from the Spanish Ministerio de Econom\'ia y Competitividad through the grant ESP2017-83197.
YKS acknowledges support from the National Research Foundation of Korea (NRF) grant funded by the Ministry of Science, ICT \& Future Planning (NRF-2019R1C1C1010279).
This work has received financial support from the European Union's Horizon 2020 Research and Innovation programme under the Marie Sklodowska-Curie grant agreement number 734374 - Project acronym: LACEGAL

\section*{Data Availability}
The data underlying this article will be shared on reasonable request to the corresponding author.





\appendix

\section{Selected Light profile fits}\label{sec:Appendix_A}
This section gives an overview of the fits to each galaxy in the sample, showing the white-light image, the best fit model and the residual image from left to right in the top row, and the models for the disc, bulge and lens component in the bottom row. Where a PSF component was included in the fit, it is not shown in these images. All images for the same galaxy are plotted to the same flux scale to allow a direct comparison of the strength of any features visible in the residual images. In general, the fits appear relatively good, with the residual light representing mainly faint asymmetric features within the galaxy such as remnant spiral arms or dust lanes within the disc.

On the right of these figures is the one-dimensional light profile fit for that galaxy. The black points show the light profile from the white-light image from a slit of width 1\arcsec\ along the major axis, and the blue, red, green and light blue lines show the light profile fits for the disc, bulge, lens and PSF components (where applicable). The purple line shows the combination of these components, and the grey dashed line gives the sky background level. The darker grey box represents the seeing FWHM, which is quantified on the bottom left of each plot.

\begin{figure*}
 \includegraphics[width=0.9\linewidth]{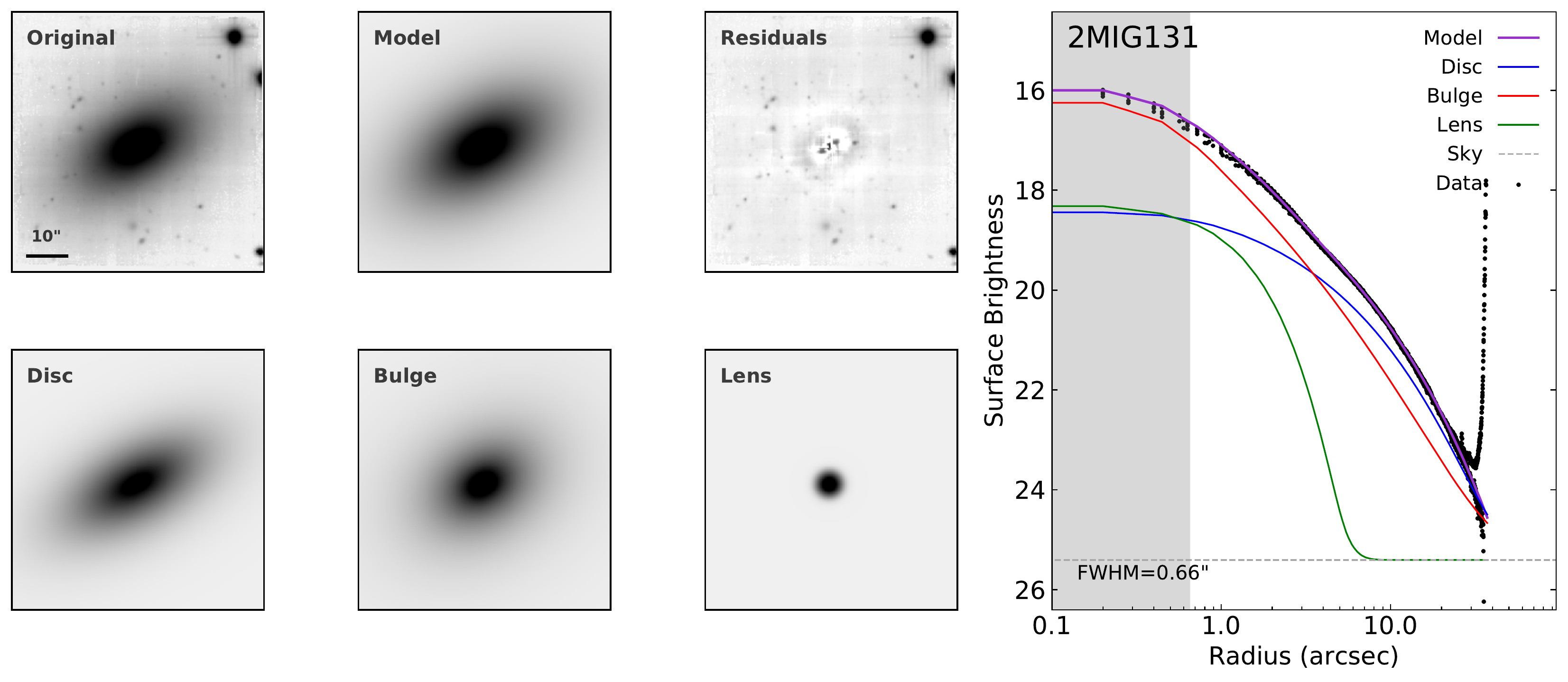}
 \includegraphics[width=0.9\linewidth]{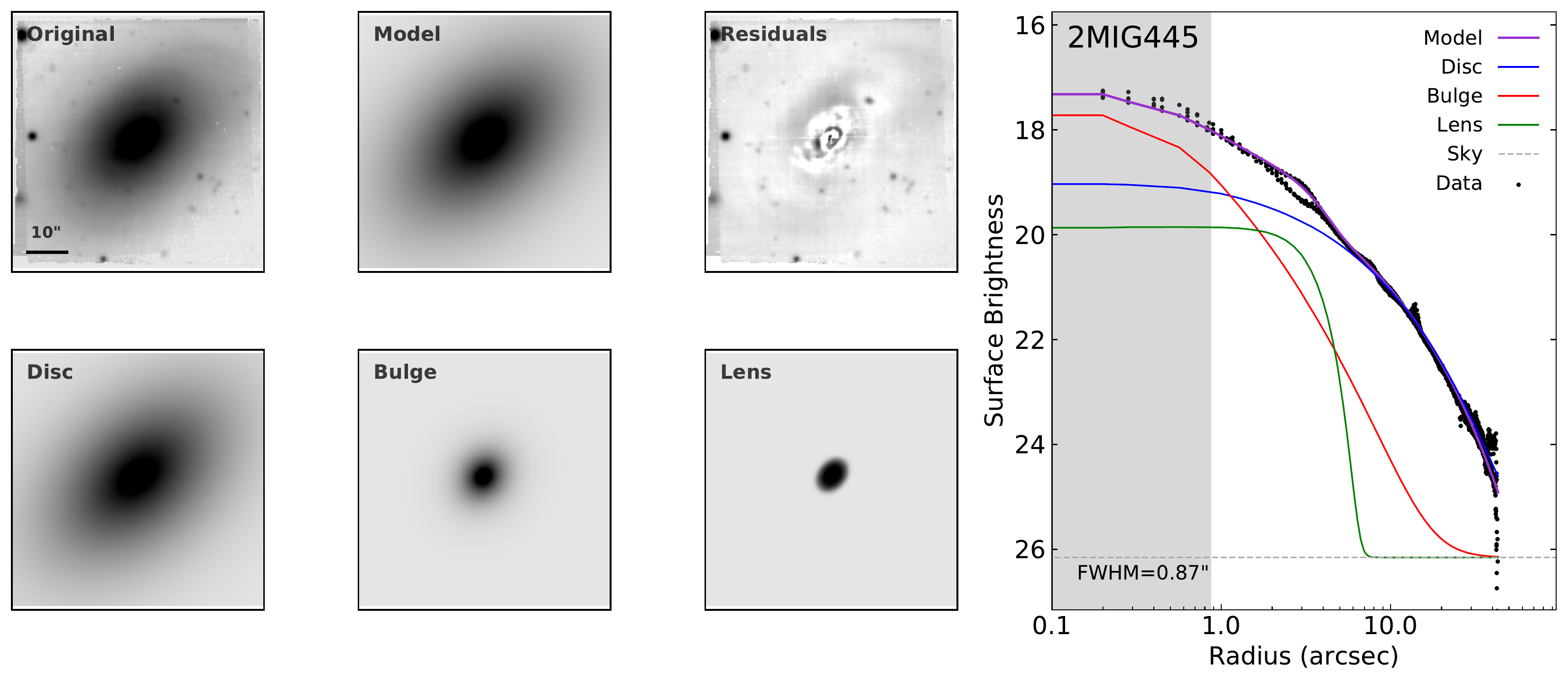}
  \caption{The fits to 2MIG~131 and 2MIG~445, showing in the top row the white-light image, the best fit model and the residual image, and below are the models for bulge, disc and lens components. The images of the PSF components are not included. On the right of these figures is the 1-dimensional light profile along the major axis of each galaxy, along with the light profiles of each component included in the model and the best fit model. The sky background is given by the horizontal dashed line, and the grey box represents the seeing FWHM measured form the datacube, with the numerical value at the bottom-left of the plot. The additional noise and sharp cut-off at high radii indicate the measurements from close to the edge of the FOV, where the S/N is lower due to fewer exposures covering that region as a result of the dithering and rotations between exposures.}
 \label{fig:fits_appendix1}
\end{figure*}
\begin{figure*}
 \includegraphics[width=0.9\linewidth]{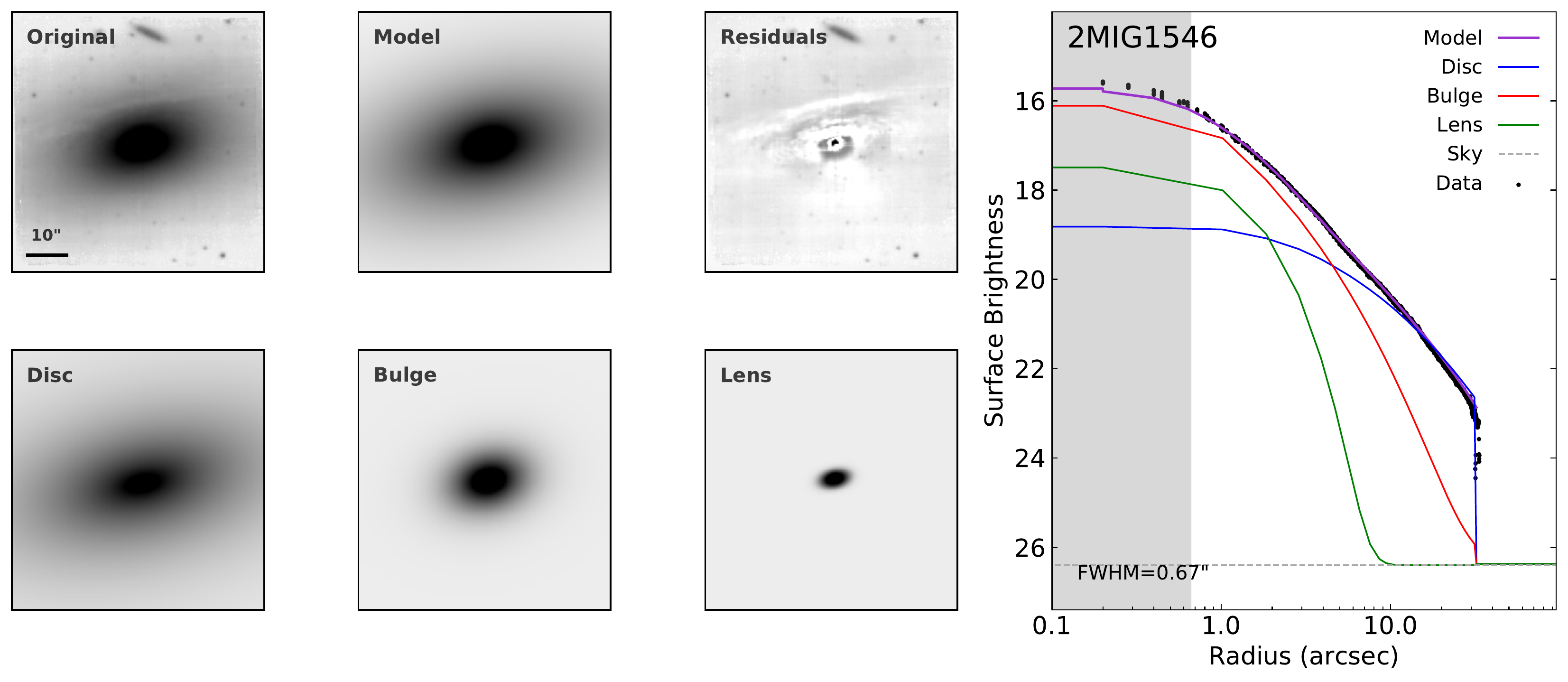}
 \includegraphics[width=0.9\linewidth]{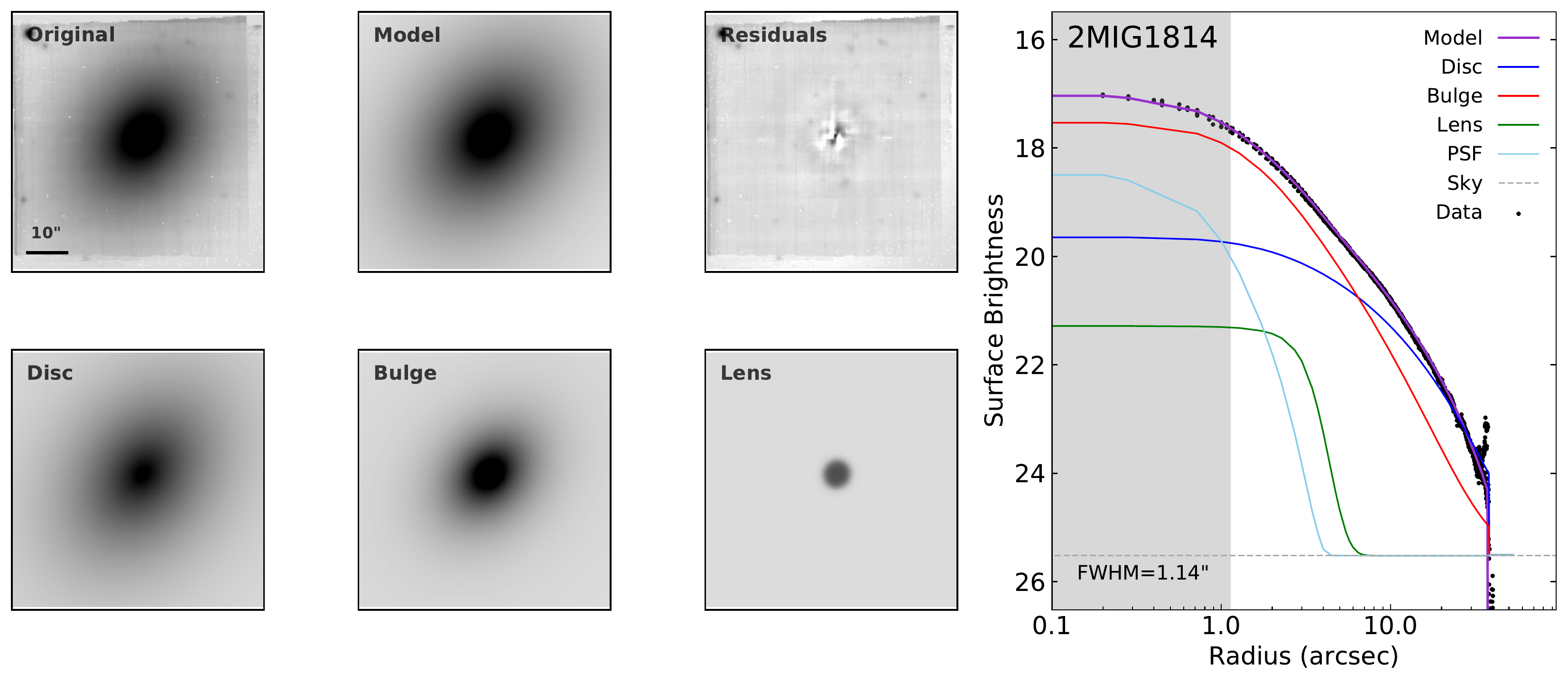}
  \includegraphics[width=0.9\linewidth]{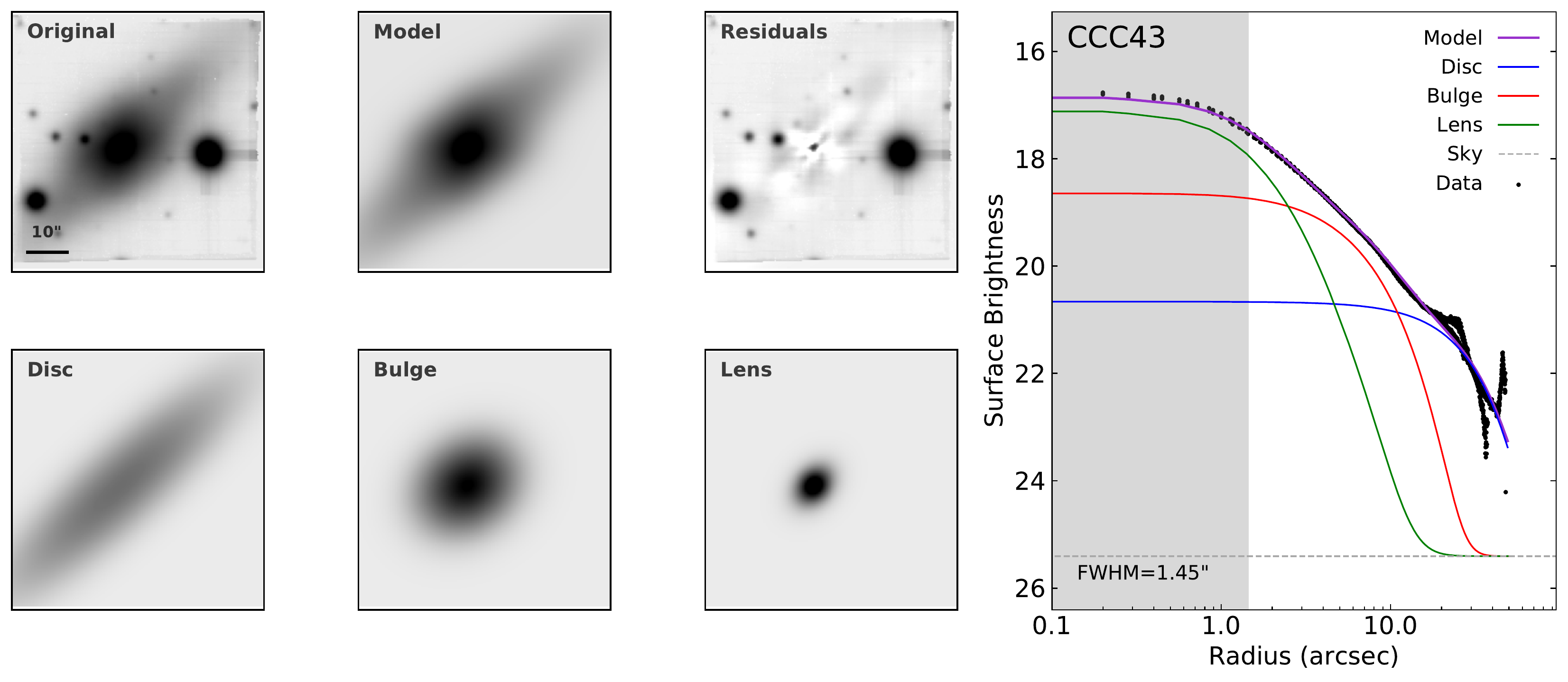}
   \caption{As for Fig.~\ref{fig:fits_appendix1}, but for  2MIG~1546, 2MIG~1814, and CCC~43. }
 \label{fig:fits_appendix2}
\end{figure*}
\begin{figure*}
  \includegraphics[width=0.9\linewidth]{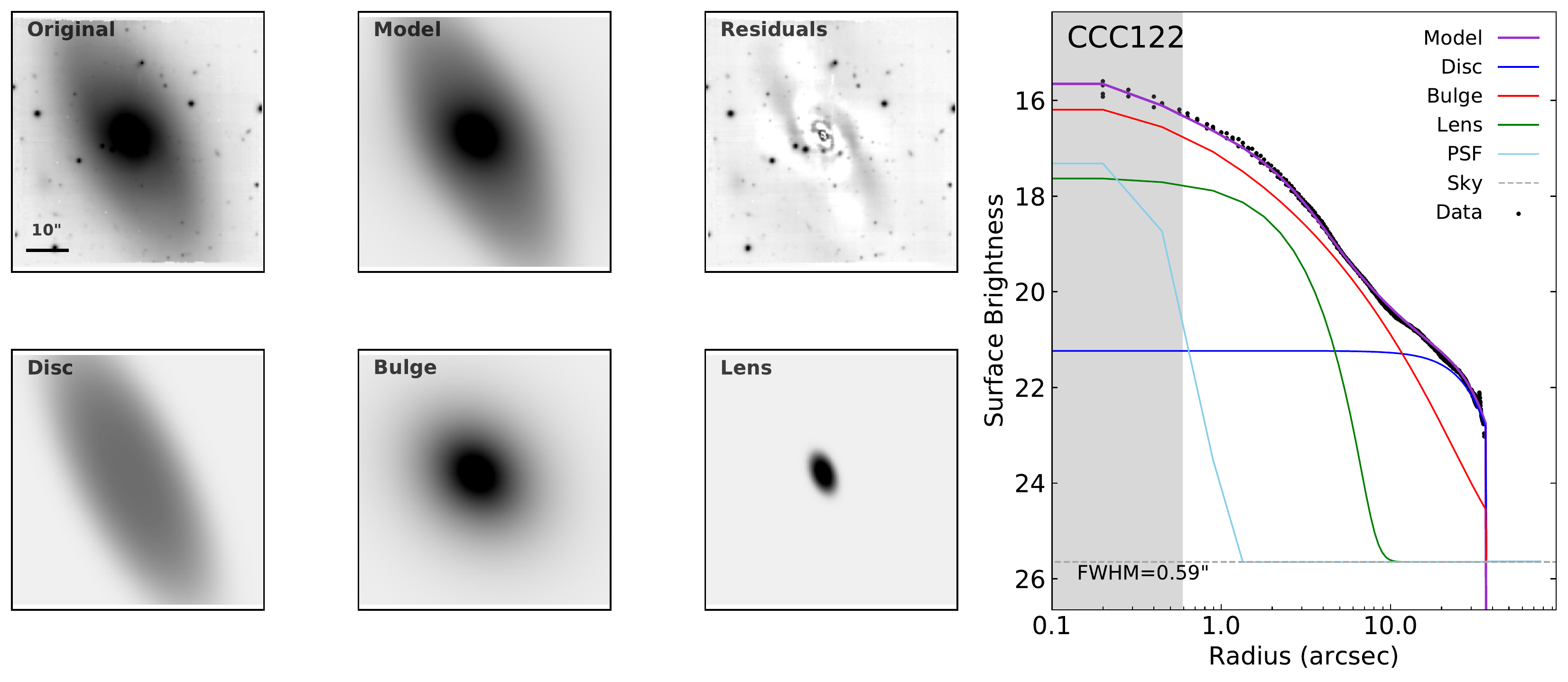}
  \includegraphics[width=0.9\linewidth]{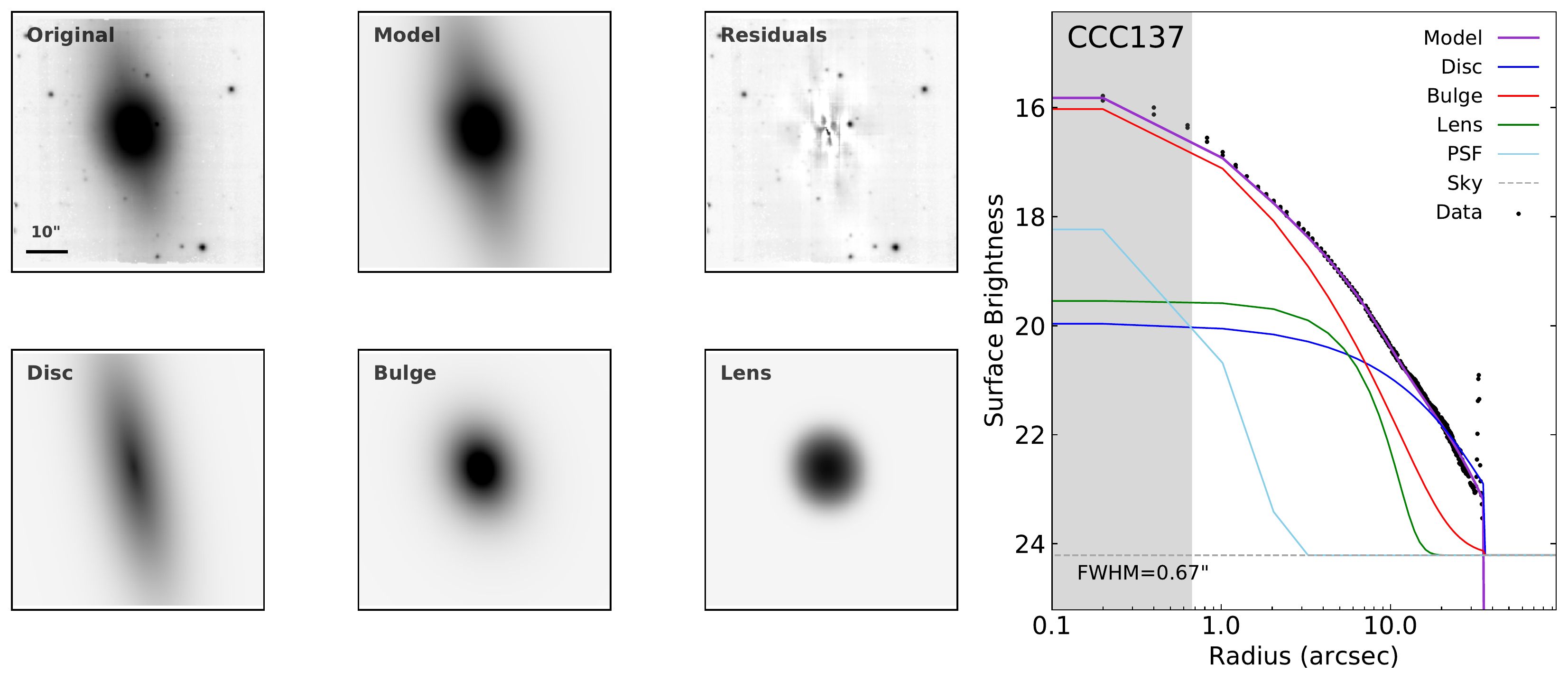}
  \includegraphics[width=0.9\linewidth]{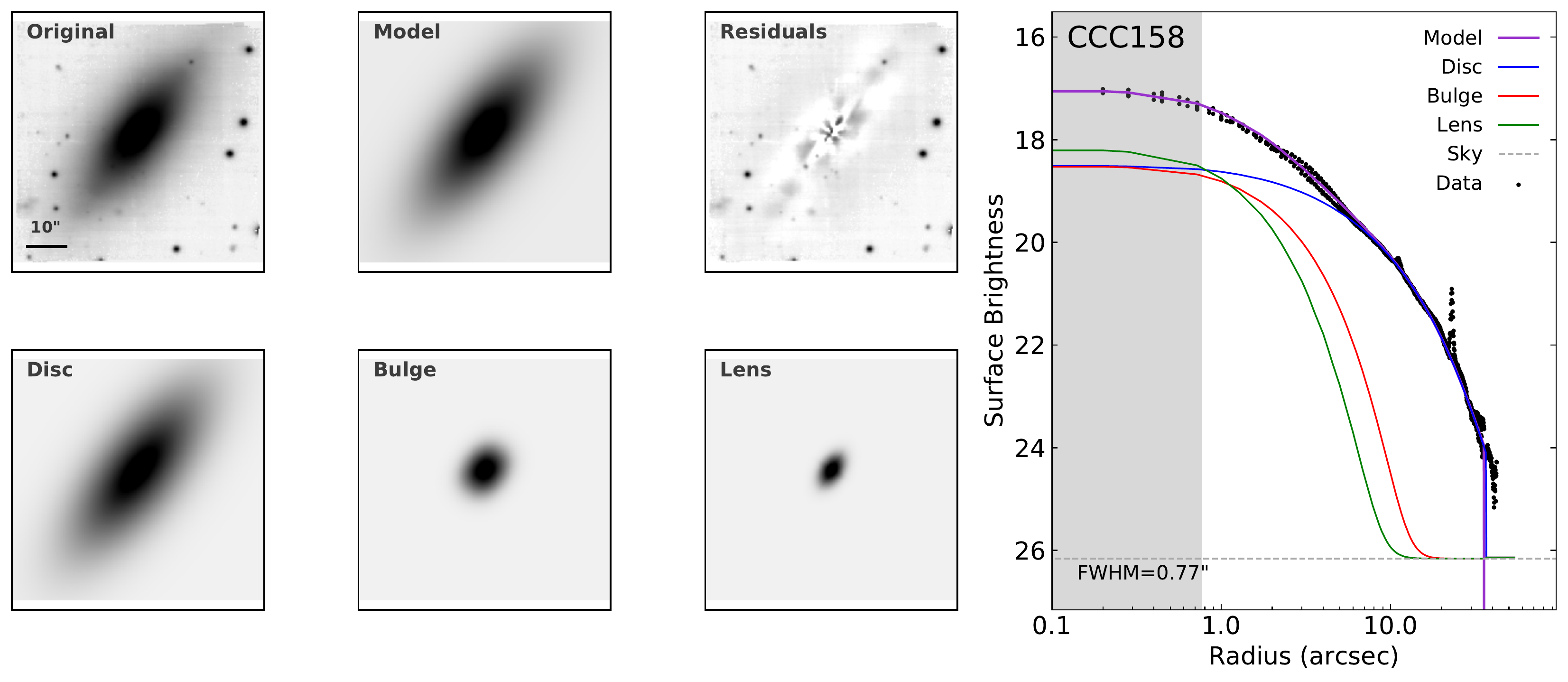}
 \caption{As for Fig.~\ref{fig:fits_appendix1}, but for CCC~122, CCC~137 and CCC~158.}
 \label{fig:fits_appendix3}
\end{figure*}


\section{2-Component Light profile fits}\label{sec:Appendix_B}
This section presents the fit to each galaxy using a simple double S\'ersic profile to represent only the bulge and disc components. In the top row are the white-light and residual images, and below the one-dimensional light profile is plotted in the same way as in Appendix~\ref{sec:Appendix_A}. All images have been scaled to the same greyscale as in Appendix~\ref{sec:Appendix_A} to allow for a direct comparison between the 2 and 3 or 4 component models.. One can see that the fits to the one-dimensional light profiles are relatively good in some cases, but in general the residual images clearly show more artefacts. In particular, one can see a bright core (indicating under subtraction of the model in that region) surrounded by alternating dark and bright rings, which are a characteristic signature that an additional component is required to improve the fit. These results are reminiscent of \citet{Laurikainen_2009}, who found that better results were obtained when fitting 2-dimensional images of their sample of S0s over a one-dimensional light profile in cases where the galaxy contains more than simply a bulge and disc.

\begin{figure*}
 \includegraphics[width=0.4\linewidth]{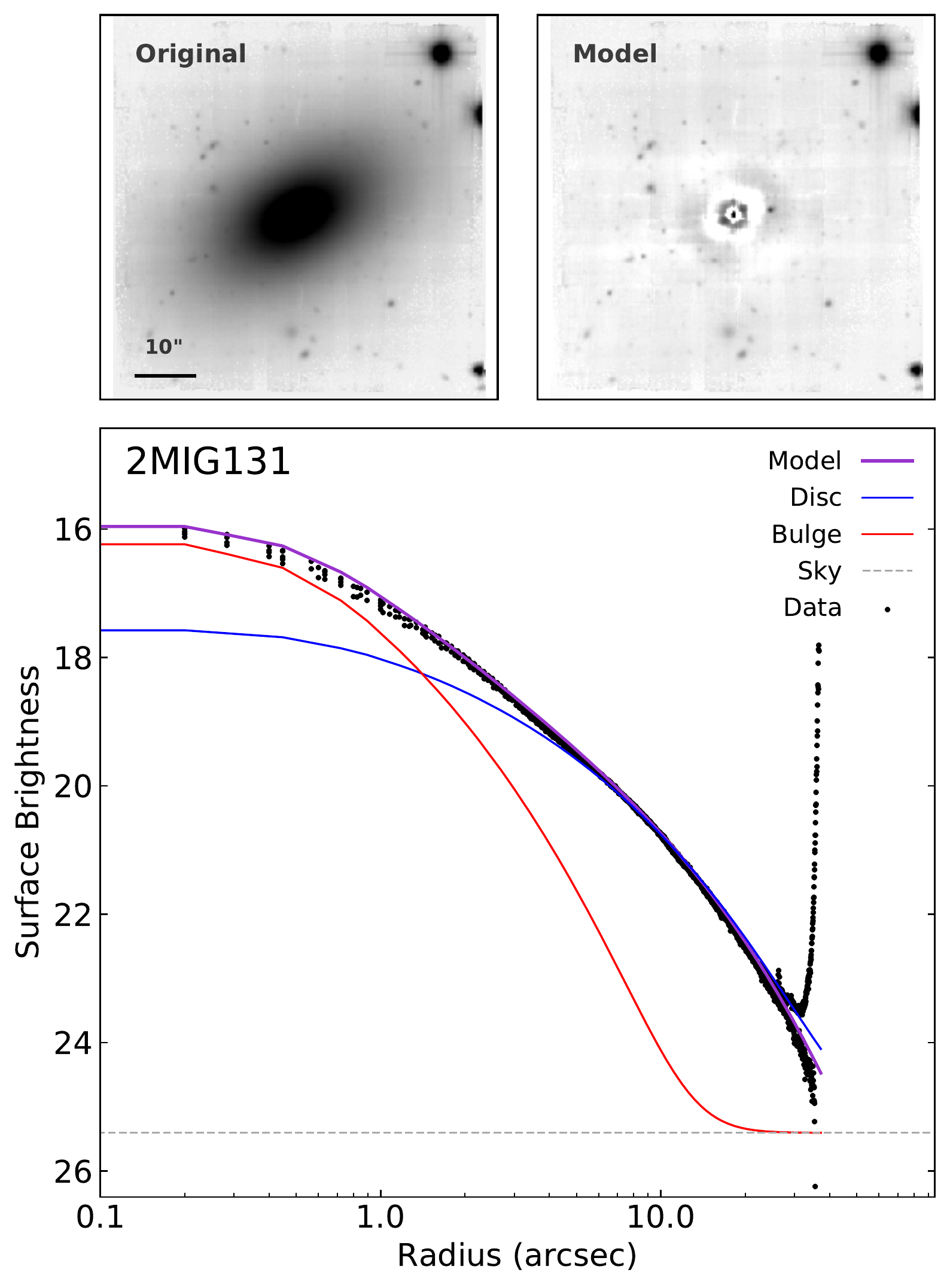}
 \includegraphics[width=0.4\linewidth]{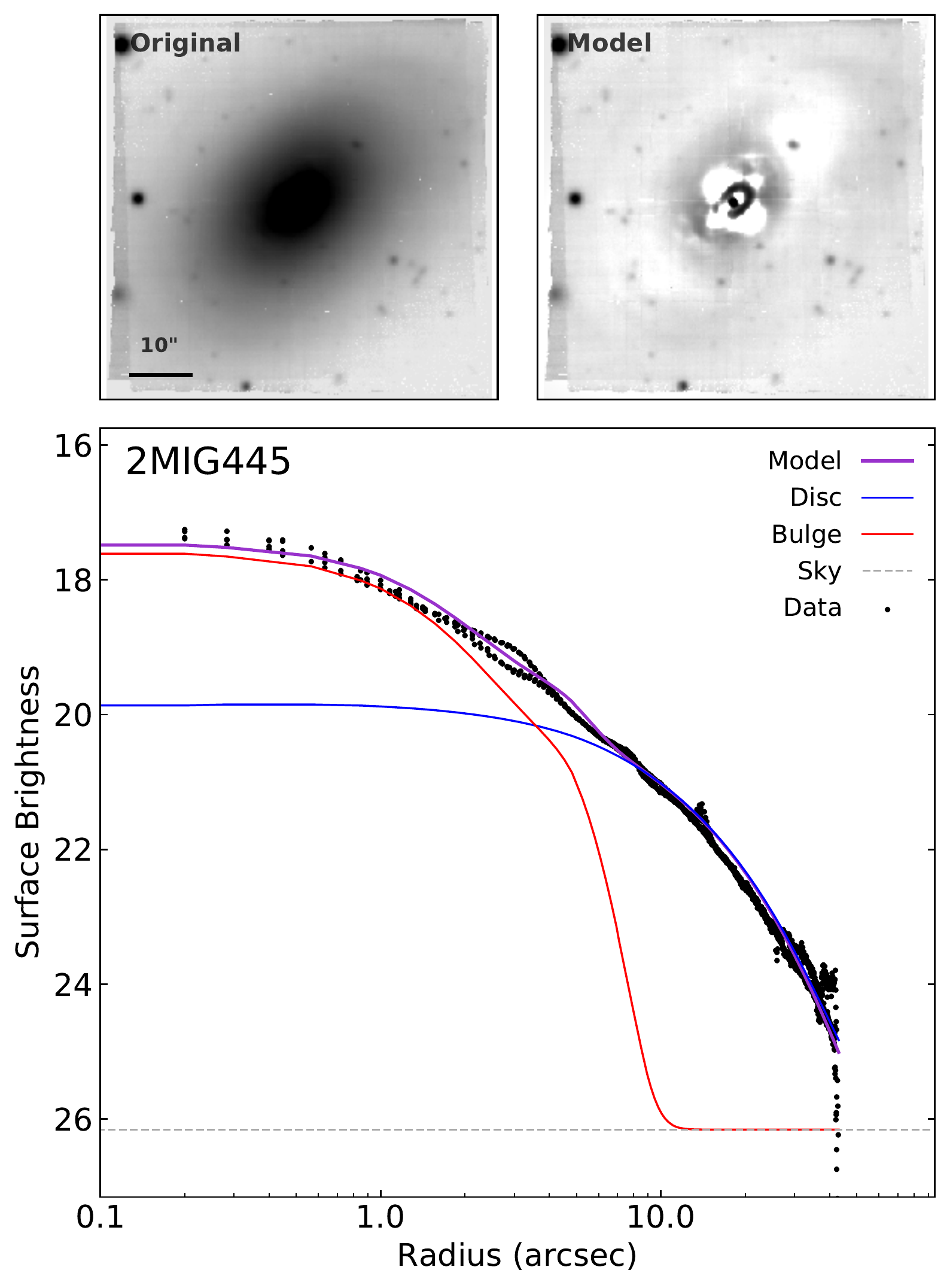}
 \includegraphics[width=0.4\linewidth]{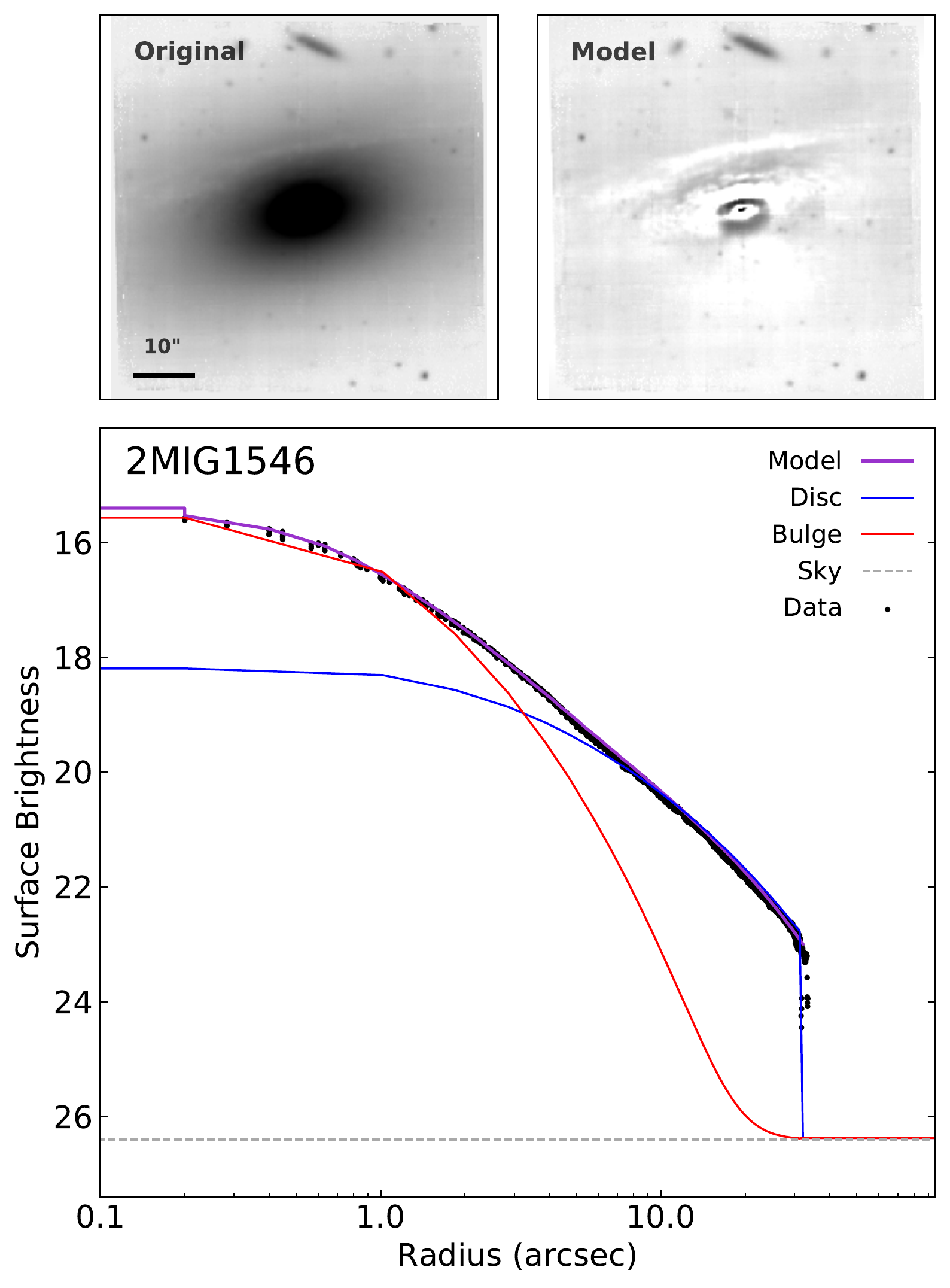}
 \includegraphics[width=0.4\linewidth]{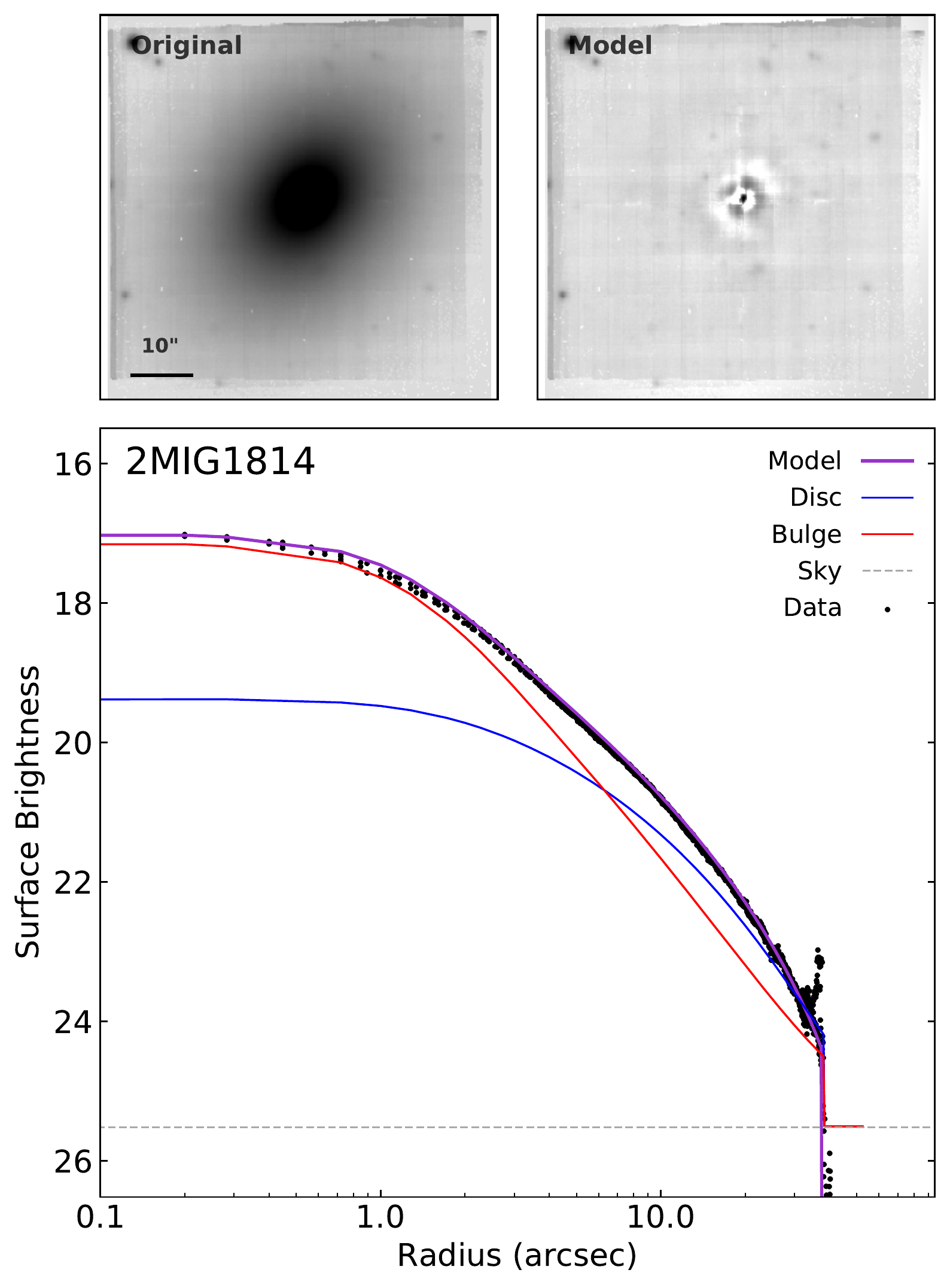}
  \caption{Double S\'ersic fits to the isolated galaxy sample, showing in each case the white-light image (top left), the residual image (top right) and the one-dimensional light profile fit (bottom).}
 \label{fig:fits_appendixB1}
\end{figure*}
\begin{figure*}
 \includegraphics[width=0.4\linewidth]{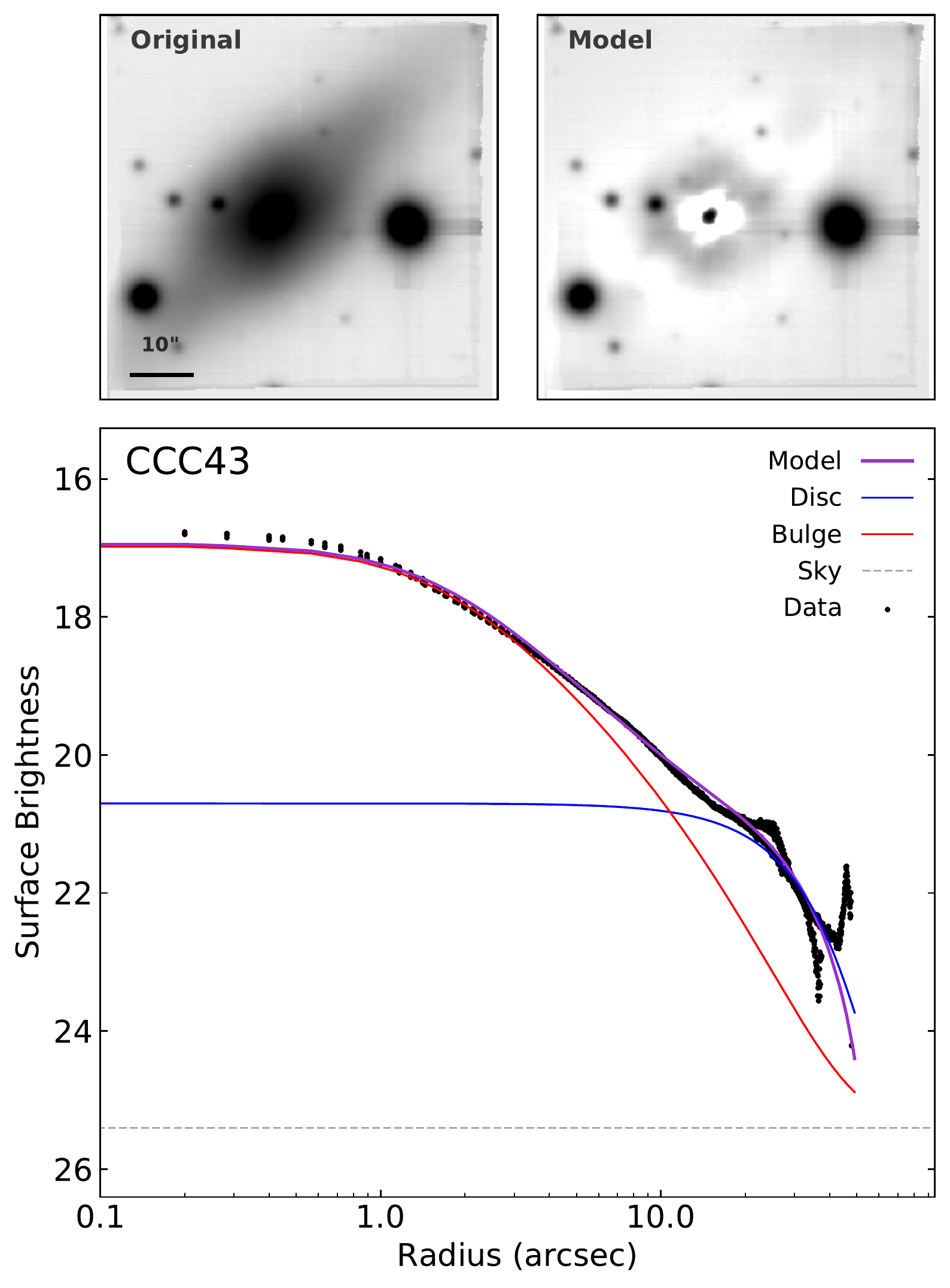}
 \includegraphics[width=0.4\linewidth]{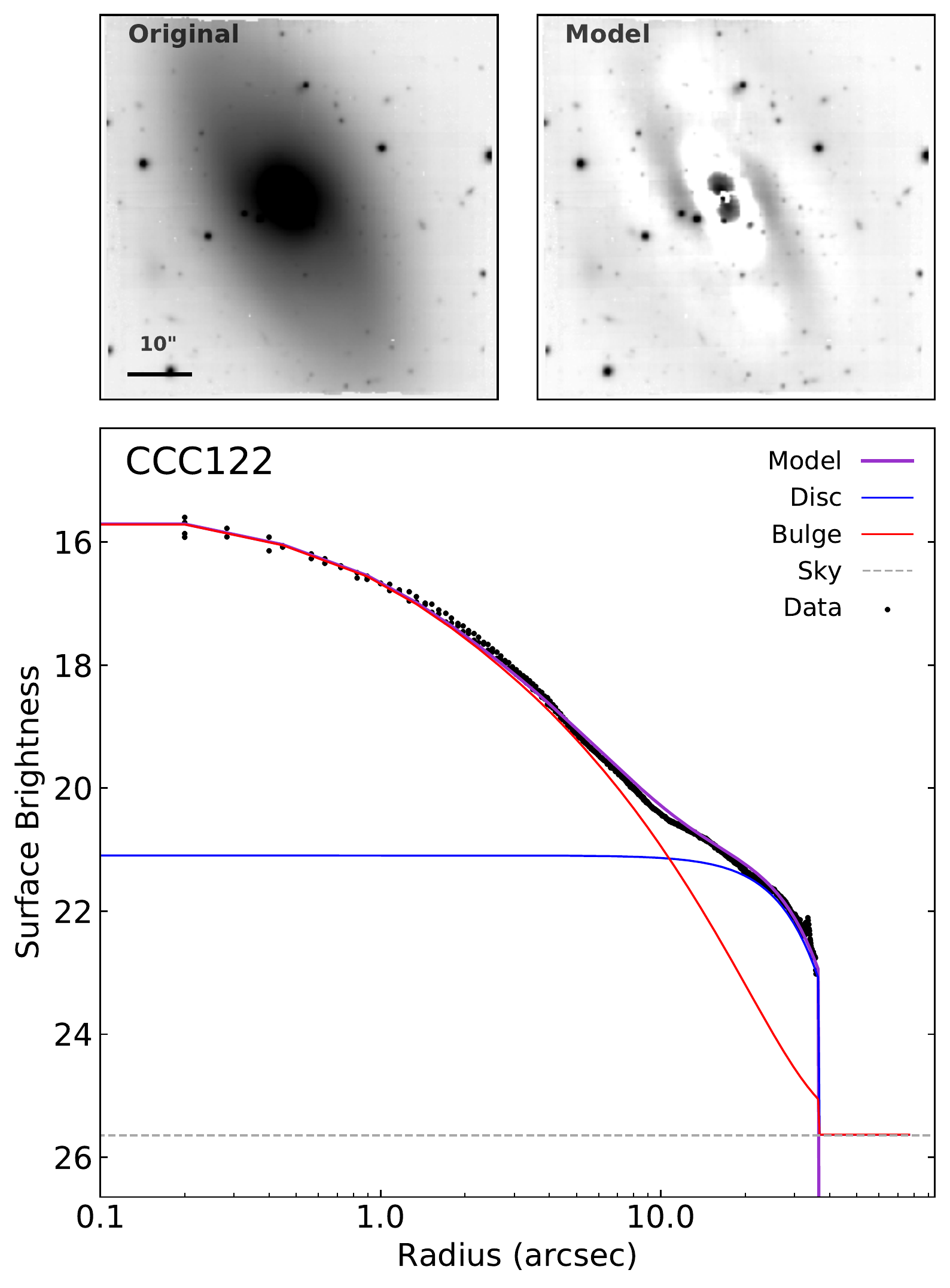}
 \includegraphics[width=0.4\linewidth]{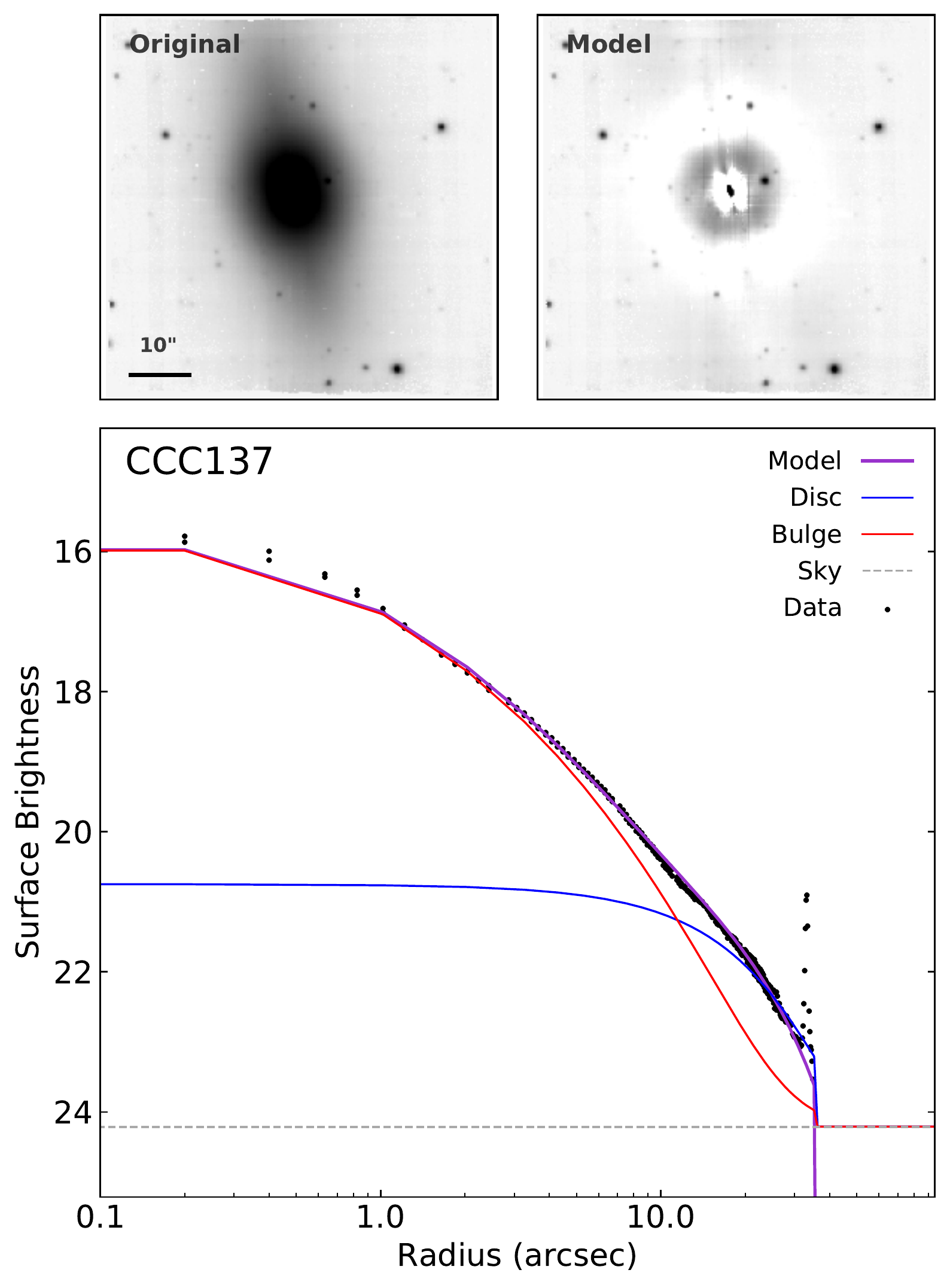}
 \includegraphics[width=0.4\linewidth]{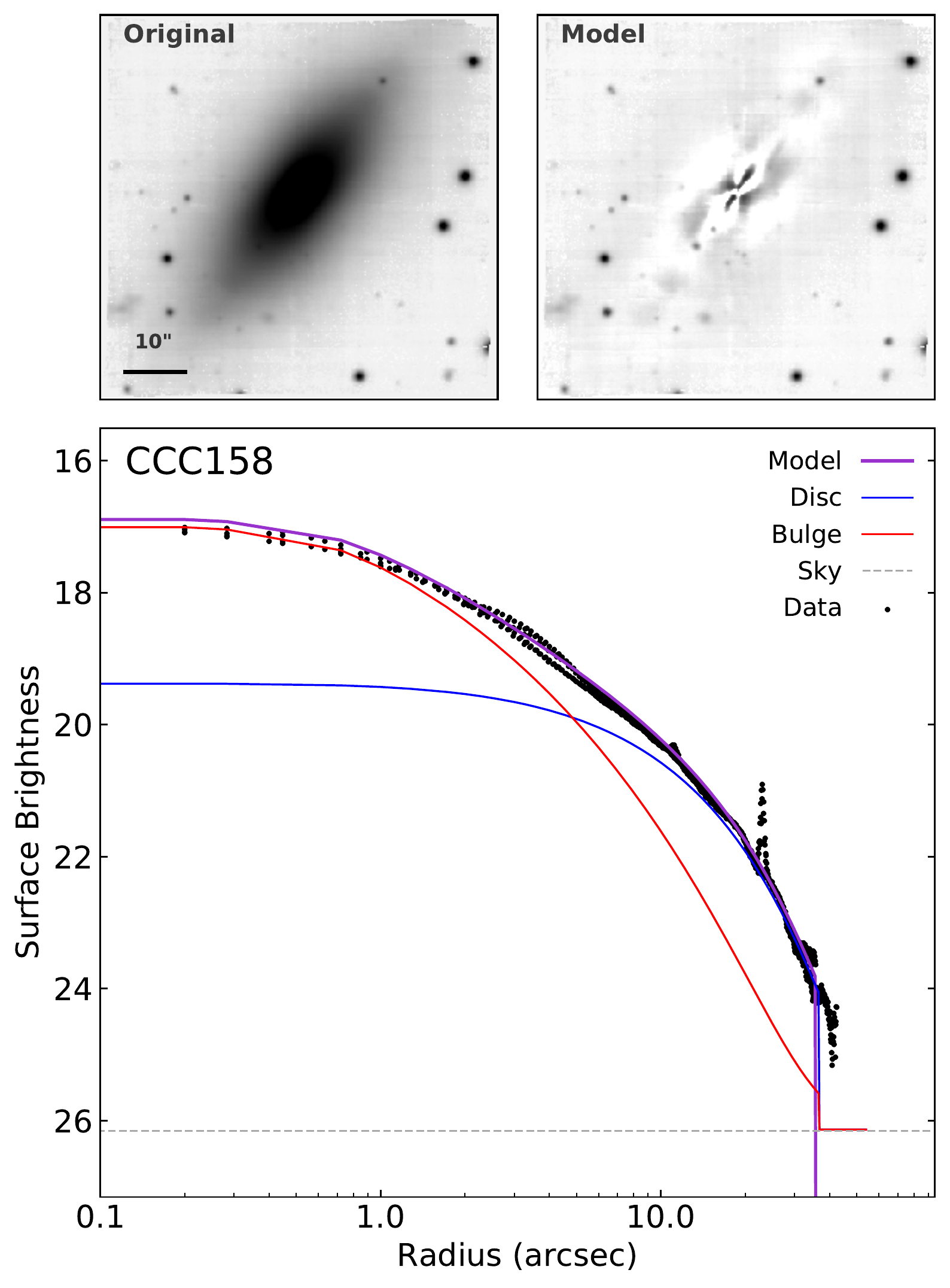}
  \caption{As for Fig.~\ref{fig:fits_appendixB1}, but for the Cluster galaxy sample.}
 \label{fig:fits_appendixB2}
\end{figure*}


\bsp	
\label{lastpage}
\end{document}